\documentclass[a4paper,12pt,leqno]{article}
\usepackage{graphicx}
\usepackage{amsmath}
\usepackage{amssymb}
\usepackage{amstext}
\usepackage{amsxtra}
\usepackage[mathscr]{eucal}
\usepackage{float}

\makeatletter
\@addtoreset{equation}{section}
\renewcommand\theequation
  {\ifnum \c@section>\z@ \thesection.\fi \@arabic\c@equation}
\makeatother

\newcommand{\II}{I\kern -1mm I}
\newcommand{\III}{I\kern -1mm I\kern -1mm I}

\newcommand{\sh}{\kern 0.5mm{\rm sh}}

\title{\bf Function Based Nonlinear Least Squares and Application to Jelinski--Moranda Software Reliability Model}
\date{}

\author{
\medskip
Jingwei Liu \footnote{Corresponding author. jwliu@buaa.edu.cn (J.W Liu)}\\
School of Mathematics and System Sciences,\\
Beihang University,\\
LMIB of the Ministry of Education, \\
Beijing, P.R China, 100191.\\
\medskip  
\\
Meizhi Xu\\
Department of Mathematics,\\
Tsinghua University,\\
Beijing, 100084, P.R China.
}

\begin{document}
\maketitle

\begin{abstract}
A function based nonlinear least squares estimation (FNLSE) method is proposed and investigated in parameter estimation of Jelinski--Moranda software reliability model. FNLSE extends the potential fitting functions of traditional least squares estimation (LSE), and takes the logarithm transformed nonlinear least squares estimation (LogLSE) as a special case. A novel power transformation function based nonlinear least squares estimation (powLSE) is proposed and applied to the parameter estimation of Jelinski--Moranda model. Solved with Newton-Raphson method, Both  LogLSE and powLSE of Jelinski--Moranda models are applied to the mean time between failures (MTBF) predications on six standard software failure time data sets. The experimental results demonstrate the effectiveness of powLSE with optimal power index compared to the classical least--squares estimation (LSE), maximum likelihood estimation (MLE) and LogLSE in terms of  recursively relative error (RE) index and Braun statistic index.


\end{abstract}

\section{Introduction}

Failure time prediction is a key problem in software reliability, which is defined as ``the probability of failure--free operation of a computer program for a specified time in a specified environment''(Musa, {\em et al}, 1987). It takes an important role in software design and software safety, especially in the development of spacecrafts, marine vessels, advanced weapons, automatic control {\em etc}.
As for the mean time between failure (MTBF) prediction, two main  methods are widely investigated in software reliability models, the time--independent model and time--dependent model.

For the time--independent model, Jelinski--Moranda model is the milestone in software reliability  to describe the MTBF of software reliability growth, with the assumption that the times between two failures are independent, the maximum likelihood estimation (MLE) and least squares estimation (LSE) are employed to estimate the model parameters (Jelinski,{\em et al}, 1972; Lyu, 1996; Cai, 1998a; Huang, 2002). However, Jelinski--Moranda model has poor prediction accuracy, Littlewood,{\em et al} (1987) owed the cause to the use of the inference with MLE, and brought forward a  Bayesian Jelinski--Moranda (BJM) model. Also, there are many models of software reliability growth to model MTBF and mean time to failure (MTTF) in the recent 3 decades (Lyu, 1996).

However, the time--dependent models take an opposite assumption that the  times between two failures are dependent, and the recent times between failures are employed to forecast the future failure time. A lot of time series based methods are developed to address the time dependent software reliability models, for example, Bayesian approach (Pham,{\em et al}, 2000; Bai,{\em et al}, 2005 ), neural network  (NN) approach (Cai, {\em et al}, 2001; Su, {\em et al}, 2007 ), support vector machine (SVM) approach (Hong and Pai, 2006a, 2006b ) and wavelet neural network (WNN) (Raj Kiran {\em et al}, 2007; Vinay Kumar  {\em et al}, 2008), {\em etc}. And,  Raj Kiran {\em et al} (2007) and Vinay Kumar  {\em et al} (2008) demonstrate that WNN is superior than many other time series based methods, for example, multilayer perceptron (MLP),radial basis function network(RBFN),multiple linear regression (MLR), dynamic evolving neuro-fuzzy inference system(DENFIS) and support vector machine(SVM), etc. These models treat themselves as black--box to predict MTBF, while time--independent models, for example Jelinski--Moranda model,  have more instinctive mathematical and statistical background.

In this paper, we focus on the time-independent modeling of Jelinski--Moranda model with LSE.
As for software reliability growth models, Schafer,{\em et al} (1979) proposed the traditional LSE technique to estimate the parameters of Jelinski-Moranda model, Cai (1998b) discussed two LSE  methods, least squares method I and least squares method II. Musa ,{\em et al} (1987) proposed a logarithm model of the exponential class model. Since Jelinski--Moranda model is an exponential class model, Liu ,{\em et al} (2008) derived the logarithm
nonlinear least squares estimation (LogLSE) of Jelinski-Moranda model and evaluated its performance on three failure data sets collected in Musa, {\em et al} (1987).

To extend the  LogLSE method, we develop a general function based nonlinear least squares estimation (FNLSE) method by combining the compression merits of transformation function in statistics with the weighted nonlinear least squares (WNLSE) to overcome the statistical modeling problem induced by heteroscedasticity (Goldfeld,1965 ; Hopkins,2003; Gujarati, 2004). We prove that FNLSE is a WNLSE method, and propose a power function based LSE (powLSE) to estimate
 the parameter of Jelinski--Moranda model.  The experimental results of MTBF prediction performances on six bench--mark failure time databases with LSE, MLE, LogLSE and powLSE, demonstrate the effectiveness of our novel powLSE model.

The rest of the paper is organized as follows. In section 2, the least squared method is reviewed,
and the FNLS method is developed . In section 3, Two FNLSE software reliability methods, LogLSE and powLSE, are discussed and the parameter estimation formula of Jelinski-Moranda model are derived. In section 4, the experimental simulation results of LogLSE and powLSE are compared on six software failure time data sets. And, the conclusion and discussion are given in the last section.

%-------------------------------------------------------------------------
\section{ Function based nonlinear least squares model}

\subsection{Least Squares Model}

Least squares method is a popular technique and widely used in many fields for function fit and parameter estimation. Let the $(x_{1},y_{1})$, $(x_{2},y_{2})$,$\cdots$, $(x_{n},y_{n})$ be observation data, the model to be fitted to the data be
\begin{equation}
y_{i}=f(x_{i}, \beta )+\epsilon_{i},
\end{equation}
where $\beta$ is the parameter vector, and $\epsilon_{i}$ is the error term.
The traditional nonlinear least squares estimation method ( Marquardt, 1963; Barham,{\em et al}, 1972) is to minimize
\begin{equation}
\begin{array}{l}
\displaystyle S=\sum_{i=1}^{n} (y_{i}-f(x_{i}, \beta ))^2. \\
\end{array}
\end{equation}
And, the nonlinear weighted least squares estimation method (Bj$\ddot{o}$rck, 1996) is minimizing
\begin{equation}
\begin{array}{l}
\displaystyle S=\sum_{i=1}^{n} w_{i}(y_{i}-f(x_{i}, \beta ))^2, \\
\end{array}
\end{equation}
where, $w_{i}> 0$.

Usually, in statistics theory, $\epsilon_{i}$ is assumed as independent variables of normal distribution $N(0,\sigma^2)$, where $\sigma^2$ is the variance of normal distribution. And,
 least squares method satisfies
 \[
 E(y)=f(x, \beta ).
 \]

If  $\epsilon_{i}\sim N(0,\sigma_{i}^2)$ , and $\sigma_{i}$ $(1\leq i\leq n)$ is not constant,
the phenomenon is called heteroscedasticity.

However, the error item of formula (1) may be non--normal. The maximum likelihood estimation (MLE) is not consistent and robust with non--normal and heteroscedastic even in the simple regression case (Marazzi,{\em et al} 2004). Briand {\em et al} (1992) also pointed out that heteroscedasticity strongly affects the prediction and interpretation of software engineering data sets, and in some sense it is difficult to use heteroscedasticity for prediction, because it is difficult to determine which piece of data has heteroscedasticity.

In statistical data analysis, two possible traditional strategies are widely adopted to fit data.
The first strategy is to find a proper function $z=H(y)$  to transform the observation data, then embedding the transformed observation data $(x_i,z_i),(i=1, \cdots n)$ to fit the formula (1), for example, the famous Box--Cox transformation, $z=\log(y)$, $z=\displaystyle \frac{a}{y}$, $z=\displaystyle e^{ay}$, {\em etc.} (Box \& Cox, 1964; Hopkins, 2003; He, 2004). Usually, the transformation function should possess compression observation data function.  The second strategy is to find the proper function $f$ to fit the formula (1) (Briand, {\em et al} 1992; Ramsay ,{\em et al} 2006; Gujarati, 2004).

However, especially for the Jelinski--Moranda model (Musa,{\em et al},1987), there is a strong assumption that $E(y)=f(x, \beta )$. Thus,
\[
\displaystyle S_{H}=\displaystyle \sum_{i=1}^{n} (H(y_{i})-f(x_{i}, \beta ))^2 \\
\]
conflicts the assumption of formula $E(y)=f(x_{i}, \beta )$.
Then, the two aforementioned strategies fail to modify the traditional least squares to enhance the prediction of software reliability.

Since weighted least squares is a strategy to overcome heteroscedasticity  (Goldfeld,1965; Markovi$\acute{c}$ ,{\em et al} 2009), we will combine the merits of transformation function with weighted least squares (WLS) to estimate the Jelinski--Moranda model in the view of data analysis, our motivation is to extend the range of parameter estimation methods to enhance the prediction rate.

\subsection{ Function based nonlinear least squares model}

Suppose that the observation data $(x_{1},y_{1})$, $(x_{2},y_{2})$,$\cdots$, $(x_{n},y_{n})$ satisfy
formula (1), where $x_{i}$, $y_{i}$, $f(x_{i}, \beta )$ are one-dimension vector and $\epsilon_{i}$ is error term. And, the function $y=H(x),( \forall x\in D \subseteq R) $ is 1--order differential, and $H'(x)\not\equiv 0$. According to Lagrange middle-value theorem, we obtain that, for any $x, x_0 \in D$, $\exists  \eta \in (x, x_0) \  \mbox{or}\  (x_0, x) $,
\begin{equation}
    H(x)=H(x_0)+ H'(\eta) (x-x_0).
\end{equation}

For any random variable $\xi$, put $x=\xi$ and $x_0=E\xi$, we obtain
\begin{equation}
    H(\xi)=H(E\xi)+ H'(\eta) (\xi-E\xi),
\end{equation}
where we still denote $ \eta \in (\xi, E\xi) \  \mbox{or}\  (E\xi, \xi)$. Obviously, $(\xi-E\xi)$ is the error item.
If $H(x)=x$, the formula (6) is the same model as formula (1).

For the observation data $y_{i}$ and $f(x_{i}, \beta )$ satisfying formula (1), put $\xi=y_{i}$, we obtain
\begin{equation}
    H(y_{i})=H(f(x_{i}, \beta )+ \epsilon_{i} )=\displaystyle H(f(x_{i}, \beta ))+ H'(\eta) \epsilon_{i}
\end{equation}
where $ \eta \in (f(x_{i}, \beta ), f(x_{i}, \beta )+ \epsilon_{i} ) \  \mbox{or}\  (f(x_{i}, \beta )+ \epsilon_{i}, f(x_{i}, \beta ))$. Even $\epsilon_{i}$ is normal distribution, the error item $ H'(\eta) \epsilon_{i} $ may be
non--normal or heteroscedastic.

Redesignate formula (6) as follows,
\begin{equation}
H(y_{i})= \displaystyle H(f(x_{i}, \beta ))+ \epsilon'_{i}, \\
\end{equation}
formula (7) denotes a function {\em H--family} based nonlinear least squares with heteroscedasticity and non--normality. Obviously, the above 1--dimension modeling discussion can be easily extended to high dimension case.

As Marazzi,{\em et al} (2004) pointed out that MLE is not consistent with heteroscedasticity and non--normality, to facilitate the data processing, we have the following definition according to the basic idea of least squares estimation, which is minimizing the sum of squares error items.

{\bf Definition 1.}  Let $(x_{1},y_{1})$, $(x_{2},y_{2})$,$\cdots$, $(x_{n},y_{n})$ be observation data, the model fitted to the data is  $y_{i}=f(x_{i}, \beta )+\epsilon_{i}$,
where $\beta$ ( $\beta \in \Theta $) is the parameter, and $\epsilon_{i}$ are independent variables of normal distribution $N(0,\sigma^2)$, $\sigma^2$ is the variance of normal distribution. The  function based nonlinear least squares estimation (FNLSE) method is to minimize
\begin{equation}
\begin{array}{l}
\displaystyle S_{H}=\displaystyle \sum_{i=1}^{n} (H(y_{i})-H(f(x_{i}, \beta )))^2, \\
\end{array}
\end{equation}
where $y=H(x), \forall x\in D $, is a continuous function with 1--order derivative, and
$H(x)\not\equiv C , \forall x\in D$, where $C$ is a constant. And, we also suppose that
$y_{i}, f(x_{i}, \beta ) \in D$, $i=1,\cdots, n $,  $\forall \beta \in \Theta $.

Therefore, the estimation of parameter $\beta$ in LSE and FNLSE takes two different styles,
\begin{equation}
\begin{array}{ll}
\displaystyle \hat{\beta}=\displaystyle \arg\min_{\beta \in \Theta} \sum_{i=1}^{n} (y_{i}-f(x_{i}, \beta ) )^2,\\ \displaystyle \hat{\beta}=\displaystyle \arg\min_{\beta \in \Theta} \{ \arg\min_{H \in \mathcal{H}} \sum_{i=1}^{n} (H(y_{i})-H(f(x_{i}, \beta )))^2 \}.\\
\end{array}
\end{equation}
where, $\mathcal{H}$ is the set of potential functions considered for FNLSE.
If $H(x)=\log_{a} x, (a>0, a\neq 1) $, we call FNLSE as LogLSE; If $H(x)=x^{\alpha}, (\alpha \neq 0) $, we call FNLSE as powLSE.

Note that, the condition that $\epsilon_{i}$ are independent variables of normal distribution $N(0,\sigma^2)$ is not necessary. If $\epsilon_{i}$ are independent variables of non--normal distribution or heteroscedastic,  Definition 1
still holds. In addition,  there are two other intuitional explanations of Definition 1,
\begin{itemize}
\item The fitting function of $y_{i}=f(x_{i}, \beta )$ is numerically extended to $H(y_{i})=H(f(x_{i}, \beta ))$ by  {\em H--family} functions including $H(x)=x$. Obviously, {\em H--family} function can be adopted as power functions $H=\{ x^{\alpha}, \alpha \neq 0 \}$, where $\alpha$ is power index.
\item The traditional error of $y_{i}$, $f(x_{i}, \beta )$ is treated as scaling along the line $H(x)=x$, if we scale the error between $y_{i}$ and $f(x_{i}, \beta )$ along any function $H(x)$, we obtain the Definition 1.
\end{itemize}

According to definition 1, The logarithm method in Musa (1987)([1],p355) takes the special
case of FNLS. Our model is different from the famous Box--Cox transformation ((Box \& Cox, 1964). Furthermore,
the FNLS is a special case of WNLSE. We have the following conclusion.

{\bf Theorem 1} If $y=H(x)$ $(\forall x\in D) $ is differential in D,
the FNLSE method $$ S_{H}=\displaystyle \sum_{i=1}^{n} (H(y_{i})-H(f(x_{i}, \beta )))^2 $$ is equivalent to a weighted nonlinear least squares estimation.

{\bf Proof:}

Since, $H(x)$ is continuous and differential in D, using Lagrangian middle-value theorem, there
exists a value $\xi_{i}\in  (y_{i}, f(x_{i}, \beta ))$ or $\xi_{i}\in (f(x_{i}, \beta ), y_{i})$, such that
\begin{equation}
H(y_{i})-H(f(x_{i}, \beta ))=H'(\xi_{i})(y_{i}-f(x_{i}, \beta ) ).
\end{equation}
Then,
\begin{equation}
\begin{array}{ll}
 S_{H}&=\displaystyle \sum_{i=1}^{n} (H(y_{i})-H(f(x_{i}, \beta )))^2
        =\displaystyle \sum_{i=1}^{n} (H'(\xi_{i}))^2(y_{i}-f(x_{i}, \beta ) )^2\\
\end{array}
\end{equation}

Let $r=(y_{1}-f(x_{1}, \beta ), y_{2}-f(x_{2}, \beta ),\cdots,y_{n}-f(x_{n}, \beta ) )^{T} $,
$W=\mbox{diag} (H'(\xi_{1}),H'(\xi_{2}),\cdots,$ $H'(\xi_{n}))$,
formula (8) takes the standard form of weighted nonlinear least squares.
\begin{equation}
   S_{H}=r^{T}Wr\\
\end{equation}

This is a weighted nonlinear least squares estimation. \hspace{4cm} $\square$

Furthermore, we have a sufficient condition to guarantee the criterion of FNLSE is smaller than criterion of LSE.

{\bf Theorem 2} If $y=H(x)$ $(\forall x\in D=[a,b])$ satisfies one of the following conditions that
\begin{enumerate}
\item [{1)}] $H(x)$ is continuous and differential in D and $H'(x)\not\equiv 0 ,|H'(x)|\leq 1, \forall x\in D$.
\item [{2)}] $H(x)$ satisfies the Lipschitz condition that $ |H(x_1)- H(x_2)|<L|x_1-x_2|$, $0<L<1$.
\end{enumerate}
Then, $ S_{H} \leq S$.

{\bf Proof:}

1) Since $ |H'(x)|<1$, we obtain
\begin{equation}
\begin{array}{ll}
\displaystyle S_{H}&=\displaystyle  \sum_{i=1}^{n} (H(y_{i})-H(f(x_{i}, \beta )))^2=\displaystyle  \sum_{i=1}^{n} (H'(\xi_{i}))^2(y_{i}-f(x_{i}, \beta ) )^2\\
                   &\leq \displaystyle  \sum_{i=1}^{n} (y_{i}-f(x_{i}, \beta ) )^2 =S
\end{array}
\end{equation}

2) Since $ |H(x_1)- H(x_2)|<L|x_1-x_2|$, $0<L<1$, we obtain
\begin{equation}
\begin{array}{ll}
\displaystyle S_{H}&=\displaystyle \sum_{i=1}^{n} (H(y_{i})-H(f(x_{i}, \beta )))^2<\displaystyle  \sum_{i=1}^{n} L^2(y_{i}-f(x_{i}, \beta ) )^2\\
                   &<\displaystyle  \sum_{i=1}^{n} (y_{i}-f(x_{i}, \beta ) )^2 =S
\end{array}
\end{equation}

Hence, this ends the proof of theorem 2.    \hspace{5.8cm} $\square$

Since logarithm function ($y=H(x)=\log_{a}(x)$) and power function ($y=H(x)=x^{\alpha}$) are two popular transformation function possessing compression of data, we mainly discuss these two functions though there are many function possessing compression property (Hopkins,2003; Gujarati, 2004). If all the $y_{i}$ and  $f(x_{i},\beta )$ fall into the interval of function's domain, FNLS of formula (6) develops a function based weighted least squares model. This is a trivial assumption that could be easily satisfied in real data analysis.
In the software reliability model, we only discuss the function with $x>0$.

It is easily to prove that the following transformation functions satisfy the condition of theorem 2 (1),

\begin{enumerate}
\item [{1)}] $\displaystyle y=H(x)=\log_{\alpha}(x), \forall x\in (  \frac{1}{ |\ln \alpha |}, +\infty ), \alpha >0, \alpha \neq 1$.
\item [{2)}] $\displaystyle y=H(x)= x^{\alpha }, \forall x\in ( |\alpha|^{\frac{1}{1-\alpha}}, +\infty ), \alpha \leq 1, \alpha \neq 0 $.
\item [{3)}] $\displaystyle y=H(x)= x^{\alpha }, \forall x\in (0, {(\frac{1}{|\alpha|})}^{\frac{1}{\alpha-1}}), \alpha > 1$.
\end{enumerate}

Because of the complexity of real data, it is difficult to determine the optimal power index $\alpha$ directly  by the previous formulae, the power index $\alpha$ optimization can be determined by the following formula,
\begin{equation}
\begin{array}{ll}
\displaystyle \hat{\alpha}_{opt}=\displaystyle  \arg\min_{\alpha \neq 0 } \{ \arg\min_{\beta \in \Theta} \sum_{i=1}^{n} (H(y_{i})-H(f(x_{i}, \beta )))^2  |_{H(x)=x^{\alpha}} \}.
\end{array}
\end{equation}

In fact, Peng, ,{\em et al}, (2003) has also applied the logarithm function based least estimation to least absolute deviations (LAD) estimation, it is different from the LogLSE style proposed by Liu,{\em et al}, (2008).

\section{FNLSE of Jelinski--Moranda Model}

In Musa,{\em et al} (1987) and Liu,{\em et al} (2008), the natural logarithm  $y=\ln(x)$ is utilized to discuss the LogLSE of Jelinski--Moranda model. We will extend LogLSE of Jelinski--Moranda model to the general FNLSE case in this section.

Suppose $x_{1},......,x_{n}$ are the time--between--failures in software fault detection,
Jelinski and Moranda (1972) assume that
\begin{enumerate}
\item [{1)}] The program totally contains N faults.
\item [{2)}] All the faults in the program are independent.
\item [{3)}] A detected fault is removed instantaneously without leading new faults to the software.
\item [{4)}] The fault detection rate remains constant over the intervals between fault occurrence.
In the $i$--th interval (from the $(i-1)$--th fault to the $i$--th fault ), the hazard rate of fault detection is $\lambda(x_i)=\phi (N-i+1)$.
And, $MTBF=\displaystyle \frac{1}{\phi (N-i+1)}$.
\item [{5)}] The being encountered probability of every fault is same in test phrase and real operation phrase.
\end{enumerate}

The $x_{i}'s$ are independent, and exponentially distributed random variables with expectation $\displaystyle \frac{1}{\phi (N-i+1)}$. The  probability density function of $x_{i}'s$ is

\[ p(x_{i})= \phi (N-i+1) \exp(- \phi (N-i+1) x_{i}). \]

The MLE of Jelinski--Moranda model is maximizing the likelihood

\[ L(N,\phi)=\prod_{i=1}^{n} \phi (N-i+1) \exp(- \phi (N-i+1) x_{i}).\]

Maximizing the log--likelihood of $L(N,\phi)$, the MLE solution satisfies the following equations:

\begin{equation}
\left\{
\begin{array}{ll}
\phi=\displaystyle \sum_{i=1}^{n} \frac{n}{\displaystyle N(\sum_{i=1}^{n} x_{i})-\sum_{i=1}^{n}(i-1)x_i} \\
\displaystyle \sum_{i=1}^{n} \frac{1}{N-i+1}= \displaystyle \frac{n}{\displaystyle N- (1/\sum_{i=1}^{n} x_{i}) (\sum_{i=1}^{n} (i-1)x_{i})}\\
\end{array}
\right.
\end{equation}

The standard LSE of Jelinski--Moranda model is minimizing
\begin{equation}
\begin{array}{l}
S(N,\phi)=\displaystyle \sum_{i=1}^{n} (x_{i}-\frac{1}{\phi (N-i+1)})^2
\end{array}
\end{equation}

Let $\displaystyle \frac{\partial S}{\partial N}=0$, $\displaystyle \frac{\partial S}{\partial \phi}=0$, the LSE of
$(N, \phi)$ satisfies the following formula,

\begin{equation}
\left\{
\begin{array}{ll}
\phi=\displaystyle \sum_{i=1}^{n} \frac{1}{(N-i+1)^2} / \sum_{i=1}^{n} \frac{x_i}{(N-i+1)} \\
\displaystyle (\sum_{i=1}^{n} \frac{x_i}{(N-i+1)^2})(\sum_{i=1}^{n} \frac{1}{(N-i+1)^2}) = \displaystyle (\sum_{i=1}^{n} \frac{x_i}{(N-i+1)})(\sum_{i=1}^{n} \frac{1}{(N-i+1)^3}) \\
\end{array}
\right.
\end{equation}

\subsection{LogLSE of Jelinski--Moranda Model}

The logarithm FNLSE of Jelinski--Moranda model is minimizing
\begin{equation}
\begin{array}{l}
S_{H}(N,\phi)=\displaystyle \sum_{i=1}^{n} (\log_{\alpha}( x_{i})-\log_{\alpha}(\frac{1}{\phi (N-i+1)}))^2 , \alpha>0.
\end{array}
\end{equation}

Since
\begin{equation}
\begin{array}{l}
S_{H}(N,\phi)=\displaystyle \frac{1}{\log(\alpha)} \sum_{i=1}^{n} (\log( x_{i})-\log(\frac{1}{\phi (N-i+1)}))^2.
\end{array}
\end{equation}
where $\log$ is natural logarithm function, also denoted as $\ln$.

Let $\displaystyle \frac{\partial S_{H}}{\partial N}=0$, $\displaystyle \frac{\partial S_{H}}{\partial \phi}=0$,
the FNLSE of $(N, \phi)$ satisfies the following formula as in Liu, {\em et al} (2008)

\begin{equation}
\left\{
\begin{array}{ll}
\displaystyle \sum_{i=1}^{n} \frac{\displaystyle \log{x_i}+\log(N-i+1)}{\displaystyle n}
 \sum_{i=1}^{n} \frac{\displaystyle 1}{\displaystyle N-i+1}
 =\displaystyle \sum_{i=1}^{n} \frac{\displaystyle \log{x_i}+\log(N-i+1)}{\displaystyle N-i+1}\\

\phi=\displaystyle \exp\{- \sum_{i=1}^{n} \frac{\displaystyle \log{x_i}+\log(N-i+1)}{\displaystyle n}\} \\
\end{array}
\right.
\end{equation}

From the above formula, we can obtain that changing the $a$ value of $H(x)=\log_{a} x$ does not affect the logarithm FNLSE of $(N,\phi)$.

\subsection{powLSE of Jelinski--Moranda model}

In this section, we will discuss another novel FNLSE for Jelinski--Moranda model with power function $ y=H(x)= x^{\alpha}$. Since $\log(y)=\log(x^{\alpha})=\alpha \log(x)$, power function can be explained as the logarithm function $\log(x)$ multiplied by a scale factor $\alpha $ ($\alpha \neq 0$)
for dependent variable $\log(y)$, which is also called log--log transformation (Hopkins, 2003).

The power function based nonlinear least squares estimation (powLSE) of Jelinski--Moranda  model is to minimize
\begin{equation}
\begin{array}{l}
S_{H}(N,\phi)=\displaystyle \sum_{i=1}^{n} (( x_{i})^{\alpha}-(\frac{1}{\phi (N-i+1)})^{\alpha})^2 , \alpha\neq 0.
\end{array}
\end{equation}

Let $\displaystyle \frac{\partial S_{H}}{\partial N}=0$, $\displaystyle \frac{\partial S_{H}}{\partial \phi}=0$, the FNLSE of $(N, \phi)$ satisfies the following formula

\begin{equation}
\left\{
\begin{array}{ll}
\displaystyle \sum_{i=1}^{n} \frac{(x_i)^{\alpha}}{(N-i+1)^{\alpha+1}} \sum_{i=1}^{n} (\frac{1}{N-i+1})^{2\alpha}=\displaystyle \sum_{i=1}^{n} (\frac{x_i}{N-i+1})^{\alpha} \sum_{i=1}^{n} (\frac{1}{N-i+1})^{2\alpha+1} \\

\phi^{\alpha}=\displaystyle \sum_{i=1}^{n} (\frac{1}{N-i+1})^{2\alpha} / \sum_{i=1}^{n} (\frac{x_i}{N-i+1})^{\alpha} \\
\end{array}
\right.
\end{equation}

When $\alpha=1$, the formula (23) is the traditional LSE of Jelinski--Moranda (Musa, {\em et al} 1987; Schafer, {\em et al} 1979; Cai, {\em et al} 1998; Huang, 2002).

Let
$$f(N)= \displaystyle \sum_{i=1}^{n} (\frac{x_i}{N-i+1})^{\alpha} \sum_{i=1}^{n} (\frac{1}{N-i+1})^{2\alpha+1}
-\sum_{i=1}^{n} \frac{(x_i)^{\alpha}}{(N-i+1)^{\alpha+1}} \sum_{i=1}^{n} (\frac{1}{N-i+1})^{2\alpha}$$

The value of $N$ is calculated from $f(N)=0$ using Newton-Raphson method, then the $\phi$ is
calculated with the corresponding formula.

Furthermore, in some cases, the  functions
\begin{enumerate}
\item [{1)}]  $y=\log_{\alpha}(x+K),\forall x>0, \alpha>0, K>0$
\item [{2)}]  $y=(x+K)^{\alpha}, \forall x>0, \alpha \neq 0, K>0$
\end{enumerate}
are also the optimal choices for data transformation in statistical data analysis. However the $(N,\phi)$ estimation of the two  functions are more complex, we omit the discussion of FNLSE with these two functions in this paper.

Since our FNLSE  does not modify any assumptions of NLS, the FNLSE of Jelinski--Moranda model does not alter the assumption of Jelinski--Moranda LSE model too, all of properties and conclusions about it still holds. As log transformation compresses the scales in which the variables are measured, the logarithm function and power function (in some sense of log--log transformation ) based FNLSE will be expected to have  accurate prediction performance too.

\subsection{Prediction Criterion and Optimization Strategy}

In the simulation experiments of software reliability, two criteria are involved in the term of recursive prediction, the RE (Cai, 1998) criterion and Braun statistic (Lyu, 1995) criterion.
For all the MLE,LSE, LogLSE and powLSE estimation of Jelinski--Moranda model, we constitute two
frameworks of evaluation.

\subsubsection{RE Criterion and Optimization Strategy}

Suppose the failure data set is $\{x_{1},\cdots, x_{n}\}$. Given power index $\alpha$ $(\alpha \neq 0)$and $i=4,\cdots, n$, the (i-1) failure times $\{x_{1},\cdots, x_{i-1} \}$ are applied to estimate the parameter $(\hat{N}, \hat{\phi})$  by MLE, LSE, LogLSE and powLSE respectively, and $(\hat{N}, \hat{\phi})$ are utilized to calculate the corresponding $\hat{MTBF_{i}}$.
Let TE denote the sum of relative errors in learning data $\{x_{1},\cdots, x_{i-1} \}$,

\begin{equation}
\begin{array}{ll}
TE_{i}=\displaystyle \sum_{j=1}^{i-1}\frac{|x_{j}-\hat{MTBF_{j}}|}{x_{j}} \times 100.
\end{array}
\end{equation}

The RE criterion of the one--step--ahead recursive prediction is defined as,
\begin{equation}
\begin{array}{ll}
RE_{i}=\displaystyle \frac{|x_{i}-\hat{MTBF_{i}}|}{x_{i}}\times 100
\end{array}
\end{equation}

The total TE and RE values in modeling data and predicting data are as follows,
\begin{equation}
\begin{array}{ll}
TE=\displaystyle \frac{1}{n-3} \sum_{i=3}^{n-1} TE_{i} \\
RE=\displaystyle \frac{1}{n-3} \sum_{i=4}^{n} RE_{i} \\
\end{array}
\end{equation}
where, we set $RE_{1}=RE_{2}=RE_{3}=0$, and $TE_{1}=TE_{2}=0$.

To optimize the power index of powLSE in formula (23), we adopt the following strategy:
For a given $\alpha$, the criteria in the model learning data of Jelinski--Moranda model with powLSE is denoted as $\{TE_{3}^{(\alpha)}, TE_{4}^{(\alpha)},\cdots,TE_{n-1}^{(\alpha)} \}$,
the The total TE values in modeling data are as follows,
\begin{equation}
\begin{array}{ll}
TE^{(\alpha)}=\displaystyle \frac{1}{n-3} \sum_{i=3}^{n-1} TE_{i}^{(\alpha)} \\
\end{array}
\end{equation}

The optimal index parameter $\alpha$ is defined as
\begin{equation}
\hat{\alpha}_{opt}=\arg\min_{\alpha \neq 0} {TE^{^{(\alpha)}}} .
\end{equation}
And, the final optimization prediction RE of powLSE is defined as $RE^{(\hat{\alpha}_{opt})}$ calculated by formula (26).

For MLE,LSE and LogLSE, there are no parameter selection, we directly calculate the TE criterion and one--step--ahead recursive prediction RE criterion.

\subsubsection{Braun statistic Criterion and Optimization Strategy}

For the failure data set $\{x_{1},\cdots, x_{n}\}$, the other important criterion is Braun statistic index, which is defined as,
\begin{equation}
\begin{array}{ll}
\displaystyle Braun statistic[\hat{MTBF_{i}}; i=s,\ldots,n ]=\displaystyle \frac{\displaystyle \sum_{i=s}^{n} (x_{i}-\hat{MTBF_{i}})^2}{\displaystyle \sum_{i=s}^{n} (x_{i}-\overline{x})^2} (\frac{n-s}{n-s-1})
\end{array}
\end{equation}
where, $\overline{x}=\displaystyle\frac{1}{n} \sum_{i=1}^{n} x_{i}$.

For a fix $i$ $(i=4,\cdots, n)$, the  segmentation of (i-1) failure times $\{x_{1},\cdots, x_{i-1} \}$  are applied to estimate the parameter $(\hat{N}, \hat{\phi})$  by MLE, LSE, LogLSE and powLSE respectively. Let $TBS_{i}$ denote the Braun statistic index in learning data $\{x_{1},\cdots, x_{i-1} \}$,

\begin{equation}
\begin{array}{ll}
TBS_{i}=\displaystyle \frac{\displaystyle \sum_{k=1}^{i-1} (x_{k}-\hat{MTBF_{k}})^2}{\displaystyle \sum_{i=1}^{i-1} (x_{k}-\overline{x})^2} (\frac{i-2}{i-3})
\end{array}
\end{equation}
where, $\overline{x}=\displaystyle\frac{1}{i-1} \sum_{k=1}^{i-1} x_{k}$.

According to the statistical meaning of Braun statistic index, the prediction of $x_i$ would be identical to the distribution of Braun statistic in training data. We denote $RBS_{i}$ as the one-- ahead--step predictive criterion,

\begin{equation}
\begin{array}{ll}
RBS_{i}=\displaystyle \frac{\displaystyle \sum_{k=1}^{i} (x_{k}-\hat{MTBF_{k}})^2}{\displaystyle \sum_{i=1}^{i} (x_{k}-\overline{x})^2} (\frac{i-1}{i-2})
\end{array}
\end{equation}
where, $\overline{x}=\displaystyle\frac{1}{i} \sum_{k=1}^{i} x_{k}$.

The overall TBS and RBS values in modeling data and predicting data are defined as follows,
\begin{equation}
\begin{array}{ll}
TBS=\displaystyle \frac{1}{n-3} \sum_{i=3}^{n-1} TBS_{i} \\
RBS=\displaystyle \frac{1}{n-3} \sum_{i=4}^{n} RBS_{i} \\
\end{array}
\end{equation}
where, we set $RBS_{1}=RBS_{2}=RBS_{3}=0$, and $TBS_{1}=TBS_{2}=0$.

For each powLSE with $\alpha$, we calculate the corresponding $TBS^{(\alpha)}$ and $RBS^{(\alpha)}$, and the optimal index is defined as

\begin{equation}
\hat{\alpha}_{opt}=\arg\min_{\alpha \neq 0} {TBS^{^{(\alpha)}}} .
\end{equation}
Then, the optimization prediction RBS of powLSE is defined as $RBS^{(\hat{\alpha}_{opt})}$ calculated by
the powLSE with optimal index $\alpha_{opt}$.

Since RE and Braun statistic index have different statistical meanings, the optimization results of power indexes with the two criteria should be different.

\subsubsection{Heteroscedasticity problem}

For the data fitting problem as addressed in formula (1), how to model the heteroscedasticity
simply from observation is very important, since it may provide information
for statistical modeling. The residue data and the variance of residual data can reflect the fluctuation of observation data. Straightforwardly, for any segmentation data $\{x_{1},\cdots, x_{i-1} \}$, ($4 \leq i\leq n$), the sample variances and the error item variance estimated by $MLE$,$LogLSE$ and $powLSE$ are easily calculated, if the variance index series along the segmentation fluctuate largely, the original time series would have Heteroscedasticity. We denote

\begin{equation}
\begin{array}{ll}
Variance=\displaystyle \frac{1}{m} \sum_{i=1}^{m} (x_{i}-\overline{x})^2 \\
Var_{MLE}=\displaystyle \frac{1}{m} \sum_{i=1}^{m} (x_{i}-\hat{x_{i}})^2 \\
Var_{LogLSE}=\displaystyle \frac{1}{m} \sum_{i=1}^{m} (x_{i}-\hat{x_{i}})^2 \\
Var_{powLSE}=\displaystyle \frac{1}{m} \sum_{i=1}^{m} (x_{i}-\hat{x_{i}})^2
\end{array}
\end{equation}
where $\hat{x_{i}}=\hat{MTBF_{i}}$ is the estimation item of $x_{i}$ by corresponding MLE, LogLSE and powLSE.

As prediction accuracy and heteroscedasticity are two interesting  problems in statistical modeling, we give a concise discussion. If the observation data have no heteroscedasticity, the fitting problem of formula (1) could be easily modeled by MLE or LSE, and the statistical model should be more correct, it could lead to more accurate prediction performance. Inversely, since the prediction accuracy acquires that the predictive value is utmost close to the real observable value, if the original data have heteroscedasticity, the predicted data series may also have heteroscedasticity.

\section{Numerical examples and simulation results}

\subsection{Data Description}

To evaluate the FNLSE performance of Jelinski--Moranda model, six  standard failure data sets are involved in the experiment.

The first failure data set, Naval Tactical Data System (NTDS) (Table 1), was first reported in
Jelinski and Moranda (1972) and evaluated in Pham, {\em et al} (2000) , it contains
34 failure data;

The following three data sets ( Table 2,3,4) appeared in Musa {\em et al} (1987). All of the three data sets were also
evaluated in Cai (1995) and our previous work Liu, {\em et al} (2008).

The fifth data set (Table 5) was from Musa (1979), it was also evaluated in Pai and Hong (2006).

The sixth data set, AT\&T (Table 6), was also evaluated in Pham, {\em et al} (2000).

\begin{table}[htbp]
\caption{NDTS failure data (Day)}
\label{table:1}
\renewcommand{\tabcolsep}{0.2pc} % enlarge column spacing
\renewcommand{\arraystretch}{0.9} % enlarge line spacing
\begin{center}
\begin{tabular}{rr|rr|rr|rr|rr|rr|rr|rr|rr}
\hline
$i$&$x_{i}$&$i$&$x_{i}$&$i$&$x_{i}$&$i$&$x_{i}$&$i$&$x_{i}$&$i$&$x_{i}$&$i$&$x_{i}$&$i$&$x_{i}$&$i$&$x_{i}$\\
\hline
1& 9  &5& 7& 9 & 5 & 13 & 1 & 17 & 3 &21& 11&  25& 2 &  29& 12 &33& 16\\
2& 12 &6& 2& 10& 7 & 14 & 9 & 18 & 3 &22& 33&  26& 1 &  30& 9  &34& 35\\
3& 11 &7& 5& 11& 1 & 15 & 4 & 19 & 6 &23& 7 &  27& 87&  31& 135& &\\
4& 4  &8& 8& 12& 6 & 16 & 1 & 20 & 1 &24& 91&  28& 47&  32& 258& &\\
\hline
\end{tabular}
\end{center}
\end{table}

\begin{table}[htbp]
\caption{JDM-I failure data (Year)}
\label{table:2}
\renewcommand{\tabcolsep}{0.15pc} % enlarge column spacing
\renewcommand{\arraystretch}{0.9} % enlarge line spacing
\begin{center}
\begin{tabular}{rr|rr|rr|rr|rr|rr|rr|rr|rr}
\hline
$i$&$x_{i}$&$i$&$x_{i}$&$i$&$x_{i}$&$i$&$x_{i}$&$i$&$x_{i}$&$i$&$x_{i}$&$i$&$x_{i}$&$i$&$x_{i}$&$i$&$x_{i}$\\
\hline
1 & 932 &3& 661 &5&1476&7&1358&9 &1169 &11& 142 &13& 660 &15& 361 &17 & 1046    \\
2 & 3103&4& 197 &6&155 &8& 288&10&1061 &12& 494 &14& 209 &16& 688 &  &\\
\hline
\end{tabular}
\end{center}
\end{table}

\begin{table}[htbp]
\caption{JDM-II failure data (Sec.)}
\label{table:3}
\renewcommand{\tabcolsep}{0.1pc} % enlarge column spacing
\renewcommand{\arraystretch}{0.9} % enlarge line spacing
\begin{center}
\begin{tabular}{r@{\quad}r|r@{\quad}r|r@{\quad}r|r@{\quad}r|r@{\quad}r|r@{\quad}r|r@{\quad}r|r@{\quad}r}
\hline
$i$&$x_{i}$&$i$&$x_{i}$&$i$&$x_{i}$&$i$&$x_{i}$&$i$&$x_{i}$&$i$&$x_{i}$&$i$&$x_{i}$&$i$&$x_{i}$\\
\hline
1& 10 &3& 13 &5& 15 &7& 18 &9& 22  &11& 19 &13& 32&15& 40 \\
2& 9  &4& 11 &6& 12 &8& 15 &10& 25 &12& 30 &14& 25&  &\\
\hline
\end{tabular}
\end{center}
\end{table}

\begin{table}[htbp]
\caption{JDM-III failure data (Sec.)}
\label{table:4}
\renewcommand{\tabcolsep}{0.15pc} % enlarge column spacing
\renewcommand{\arraystretch}{0.9}% enlarge line spacing
\begin{center}
\begin{tabular}{rr|rr|rr|rr|rr|rr|rr|rr}
\hline
$i$&$x_{i}$&$i$&$x_{i}$&$i$&$x_{i}$&$i$&$x_{i}$&$i$&$x_{i}$&$i$&$x_{i}$&$i$&$x_{i}$&$i$&$x_{i}$\\
\hline
1 &320   &22& 6499 &43& 2880  &64& 149606 &85& 86400 &106& 10506   &127&  3600 &148& 432000 \\
2 &1439  &23& 3124 &44& 110	  &65& 14400  &86&288000 &107& 177240  &128&144000 &149& 1411200\\
3 &9000	 &24& 51323&45& 22080 &66& 34560  &87&  320	 &108& 241487  &129& 14400 &150& 172800\\
4 &2880	 &25& 17010&46& 60654 &67& 39600  &88& 57600 &109& 143028  &130& 86400 &151& 86400 \\
5 &5700	 &26&  1890&47&	52163 &68& 334395 &89& 28800 &110& 273564  &131&110100 &152& 1123200\\
6 &21800 &27&  5400&48&	12546 &69& 296015 &90& 18000 &111& 189391  &132& 28800 &153& 1555200\\
7 &26800 &28& 62313&49&	784	  &70& 177395 &91& 88640 &112& 172800  &133& 43200 &154& 777600\\
8 &113540&29& 24826&50&	10193 &71& 214622 &92&432000 &113& 21600   &134& 57600 &155&1296000\\
9 &112137&30& 26355&51&	7841  &72& 156400 &93& 4160	 &114& 64800   &135&468000 &156& 1872000\\
10&660	 &31&	363&52&	31365 &73& 166800 &94& 3200	 &115& 302400  &136&950400 &157& 335600\\
11&2700	 &32&13989 &53&	24313 &74& 10800  &95&42800	 &116& 752188  &137&400400 &158& 921600 \\
12&28793 &33&15058 &54&	298890&75& 267000 &96&43600	 &117& 86400   &138&883800 &159& 1036800\\
13&2173	 &34&32377 &55&	1280  &76&  34513 &97&10560	 &118& 100800  &139&273600 &160& 1728000\\
14&7263	 &35&41632 &56&	22099 &77&	7680  &98&115200 &119& 19440   &140&432000 &161& 777600\\
15&10865 &36&4160  &57&	19150 &78&	37667 &99&86400	 &120& 115200  &141&864000 &162& 57600 \\
16&4230	 &37&82040 &58&	2611  &79&	11100 &100&57600 &121& 64800   &142&202600 &163& 17280 \\
17&8460	 &38&13189 &59&	39170 &80&	187200&101&	28800	&122& 3600	  &143&	203400	&   &\\
18&14805 &39&3426  &60&	55794 &81&	18000 &102&	432000	&123& 230400  &144&	277680	&   &\\
19&11844 &40&5833  &61&	42632 &82&	178200&103&	345600	&124& 583200  &145&	105000	&   &\\
20&5361	 &41&640   &62&	267600&83&	144000&104&	115200	&125& 259200  &146&	580080	&   &\\
21&6553	 &42&640   &63&	87074 &84&	639200&105&	44494	&126& 183600  &147&	4533960	&   &\\
\hline
\end{tabular}
\end{center}
\end{table}

\begin{table}[H]
\caption{JDM-IV failure data (Sec.)}
\label{table:5}
\renewcommand{\tabcolsep}{0.15pc} % enlarge column spacing
\renewcommand{\arraystretch}{0.9} % enlarge line spacing
\begin{center}
\begin{tabular}{rr|rr|rr|rr|rr|rr|rr}
\hline
$i$&$x_{i}$&$i$&$x_{i}$&$i$&$x_{i}$&$i$&$x_{i}$&$i$&$x_{i}$&$i$&$x_{i}$&$i$&$x_{i}$\\
\hline
1& 5.7683	&16&8.3499	&31&5.8944	&46&11.0129	&61&10.6604	&76&14.5569	&91&11.3667\\
2& 9.5743	&17&9.0431	&32&9.546	&47&10.8621	&62&12.4972	&77&13.3279	&92&11.3923\\
3& 9.105	&18&9.6027	&33&9.6197	&48&9.4372	&63&11.3745	&78&8.9464	&93&14.4113\\
4& 7.9655	&19&9.3736	&34&10.3852	&49&6.6644	&64&11.9158	&79&14.7824	&94&8.3333\\
5& 8.6482	&20&8.5869	&35&10.6301	&50&9.2294	&65&9.575	&80&14.8969	&95&8.0709\\
6& 9.9887	&21&8.7877	&36&8.3333	&51&8.9671	&66&10.4504	&81&12.1399	&96&12.2021\\
7& 10.1962	&22&8.7794	&37&11.315	&52&10.3534	&67&10.5866	&82&9.7981	&97&12.7831\\
8& 11.6399	&23&8.0469	&38&9.4871	&53&10.0998	&68&12.7201	&83&12.0907	&98&13.1585\\
9& 11.6275	&24&10.8459	&39&8.1391	&54&12.6078	&69&12.5982	&84&13.0977	&99&12.753\\
10& 6.4922	&25&8.7416	&40&8.6713	&55&7.1546	&70&12.0859	&85&13.368	&100&10.3533\\
11& 7.901	&26&7.5443	&41&6.4615	&56&10.0033	&71&12.2766	&86&12.7206	&101&12.4897\\
12& 10.2679	&27&8.5941	&42&6.4615	&57&9.8601	&72&11.9602	&87&14.192	&   &    \\
13& 7.6839	&28&11.0399	&43&7.6955	&58&7.8675	&73&12.0246	&88&11.3704	&   &    \\
14& 8.8905	&29&10.1196	&44&4.7005	&59&10.5757	&74&9.2873	&89&12.2021	&   &    \\
15& 9.2933	&30&10.1786	&45&10.0024	&60&10.9294	&75&12.495	&90&12.2793	&   &    \\
\hline
\end{tabular}
\end{center}
\end{table}

\begin{table}[H]
\caption{AT\& T Bell failure data (In CPU Units)}
\label{table:6}
\renewcommand{\tabcolsep}{0.15pc} % enlarge column spacing
\renewcommand{\arraystretch}{0.9} % enlarge line spacing
\begin{center}
\begin{tabular}{rr|rr|rr|rr|rr|rr|rr|rr}
\hline
$i$&$x_{i}$&$i$&$x_{i}$&$i$&$x_{i}$&$i$&$x_{i}$&$i$&$x_{i}$&$i$&$x_{i}$&$i$&$x_{i}$&$i$&$x_{i}$\\
\hline
1&5.50 &4&70.89 &7&  3.47  &10& 19.88& 13& 11.42 &16& 0.04   &19 &0.45  &22&47.6 \\
2&1.83 &5&3.94  &8&  9.96  &11& 7.81&  14& 18.94 &17& 125.67 &20 &31.61 &    &\\
3&2.75 &6&14.98 &9&  11.39 &12& 14.59& 15& 65.3  &18& 82.69  &21 &129.31&    &\\
\hline
\end{tabular}
\end{center}
\end{table}

All of the six data sets are bench--mark software failure data, they are evaluated in many references.
All of them are employed to evaluate our FNLSE of Jelinski--Moranda model.

\subsection{Experimental Results of FNLS Jelinski--Moranda Model}

In the simulations of Jelinski--Moranda model, all of the MLE, LogLSE and powLSE
are applied to the estimation of parameters in Jelinski--Moranda model, the power index ranges
in the set of $\{-2,-7/4,-3/2,-1,-3/4,-1/2,-1/4,1/4,1/2,3/4,1,5/4,3/2,7/4,2\}$, that is from -2 to 2 by step of 1/4.

For each standard failure data set $\{x_{1},\cdots, x_{n}\}$, we estimate the parameters of Jelinski--Moranda model by MLE, LogLSE and powLSE in every segmentation data of $\{x_{1},\cdots, x_{m}\}$ ($3 \leq m \leq n-1$), and calculate the variances of original data and the residual data determined by MLE, LogLSE and powLSE respectively.

\subsubsection{Experimental Results with RE criterion }

The TE criteria of the recursive training data are shown from Fig. 1  to Fig. 6. The relationship of TE criteria and the corresponding RE criterion with same index of powLSE are shown in Fig. 7 -- Fig. 12.
And, the RE values of MLE, LogLSE and powLSE with optimal index are listed in Table 7.

\begin{figure}[H]
\begin{center}
\begin{minipage}{5.5cm}
\includegraphics[width=5.5cm]{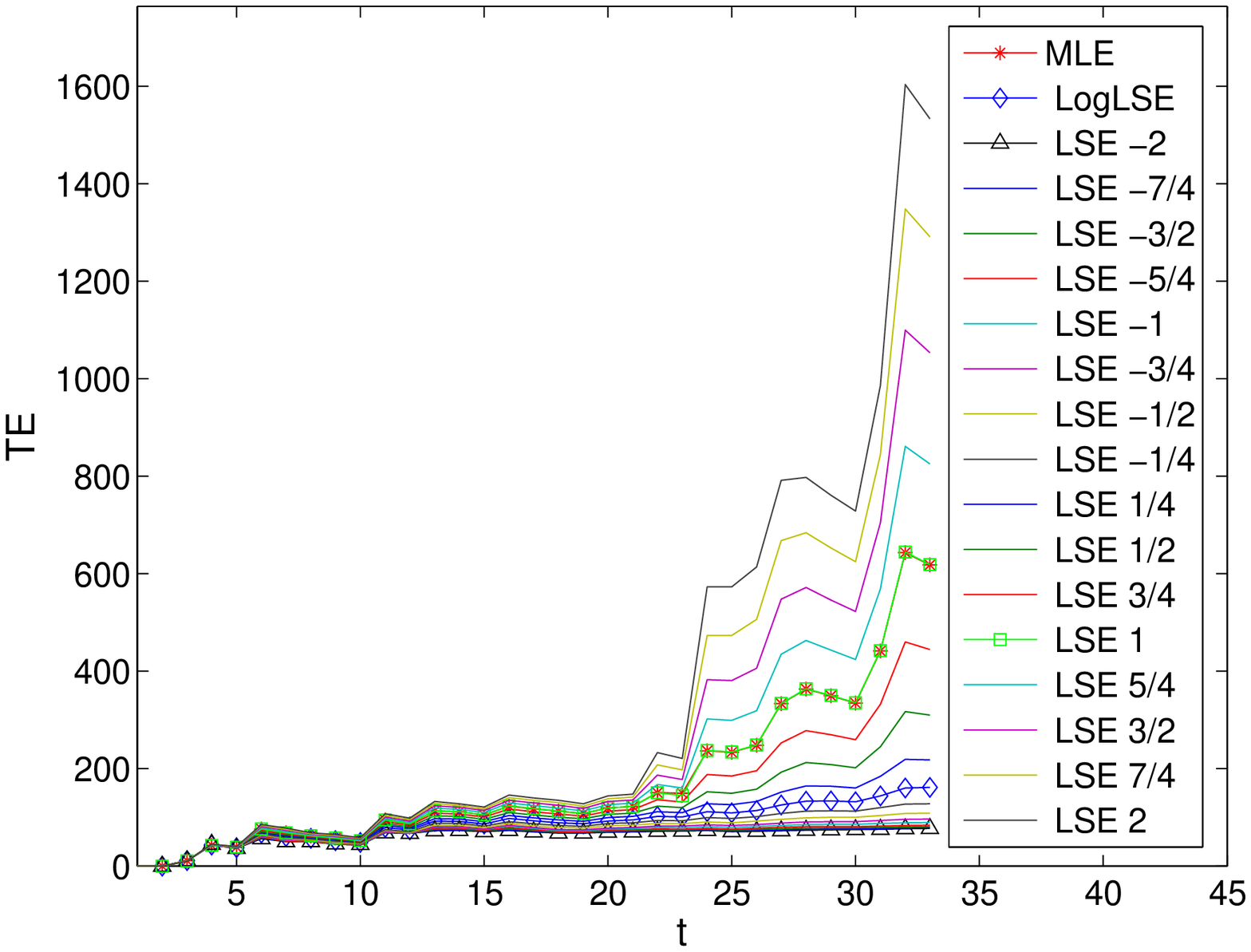}
\caption{TE values of NTDS with MLE, LogLSE and powLSE.}
\end{minipage}
\hspace{1cm}
\begin{minipage}{5.5cm}
\includegraphics[width=5.5cm]{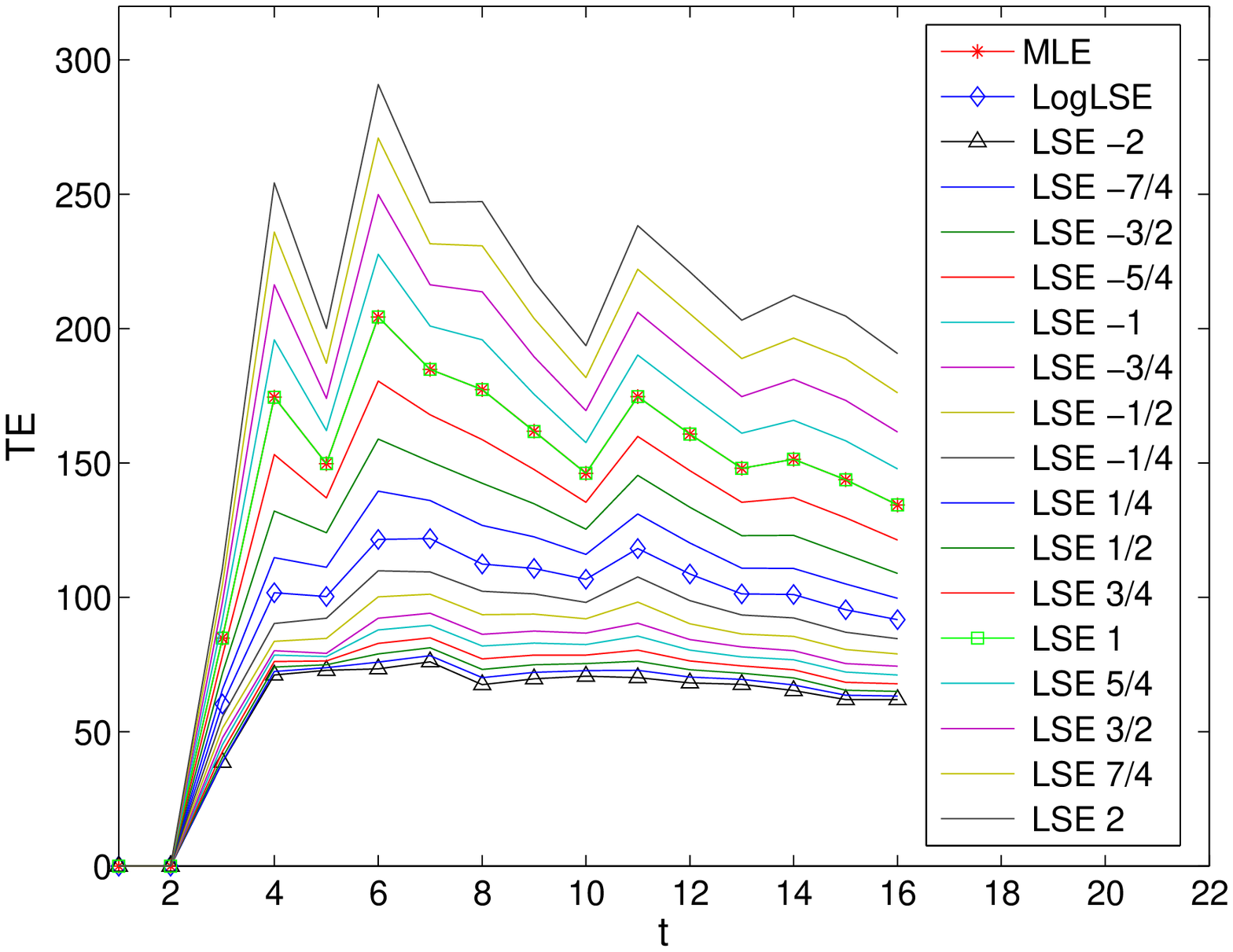}
\caption{TE values of JDM--I with MLE, LogLSE and powLSE.}
\end{minipage}
\end{center}
\end{figure}

\begin{figure}[H]
\begin{center}
\begin{minipage}{5.5cm}
\includegraphics[width=5.5cm]{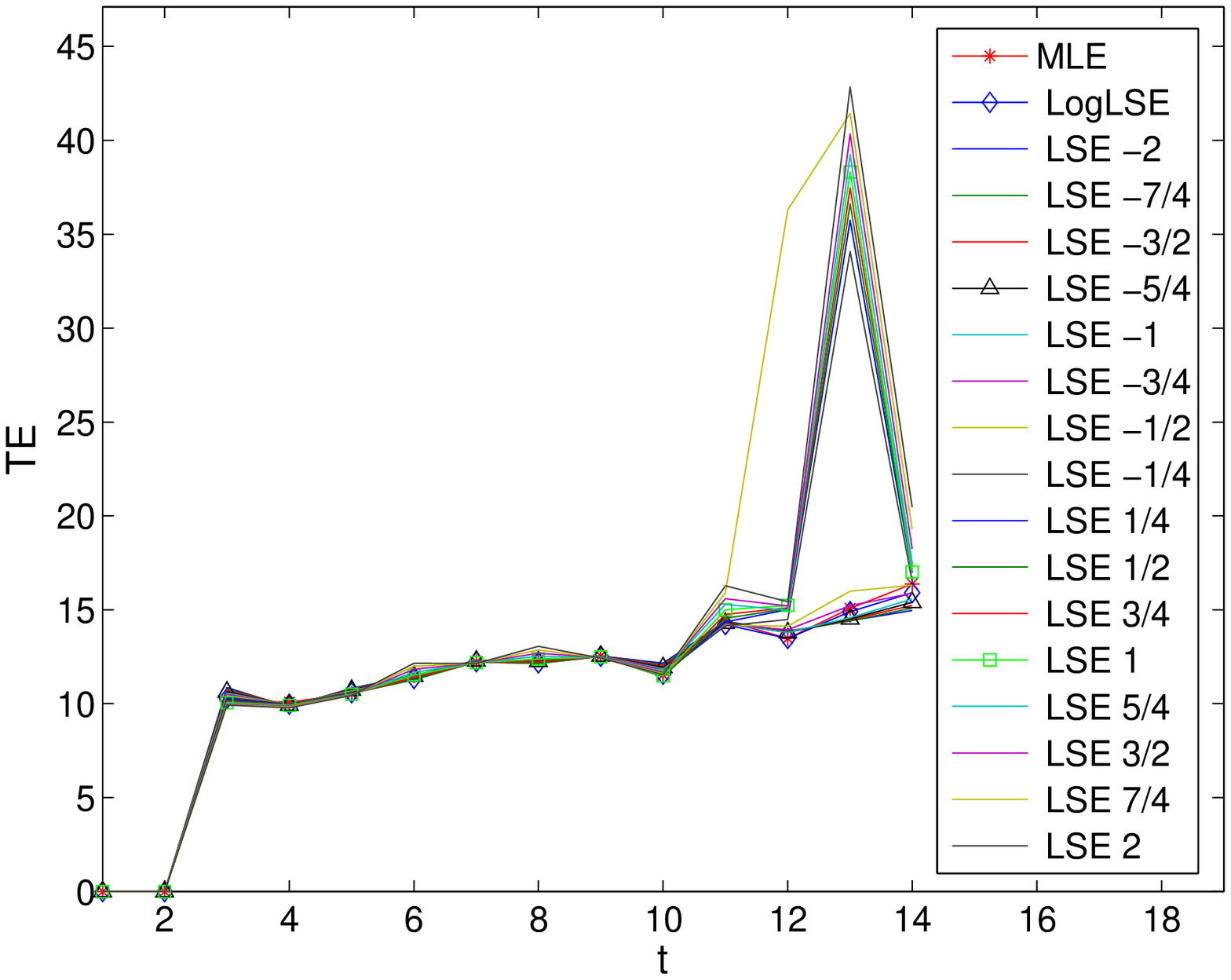}
\caption{TE values of JDM--II with MLE, LogLSE and powLSE.}
\end{minipage}
\hspace{1cm}
\begin{minipage}{5.5cm}
\includegraphics[width=5.5cm]{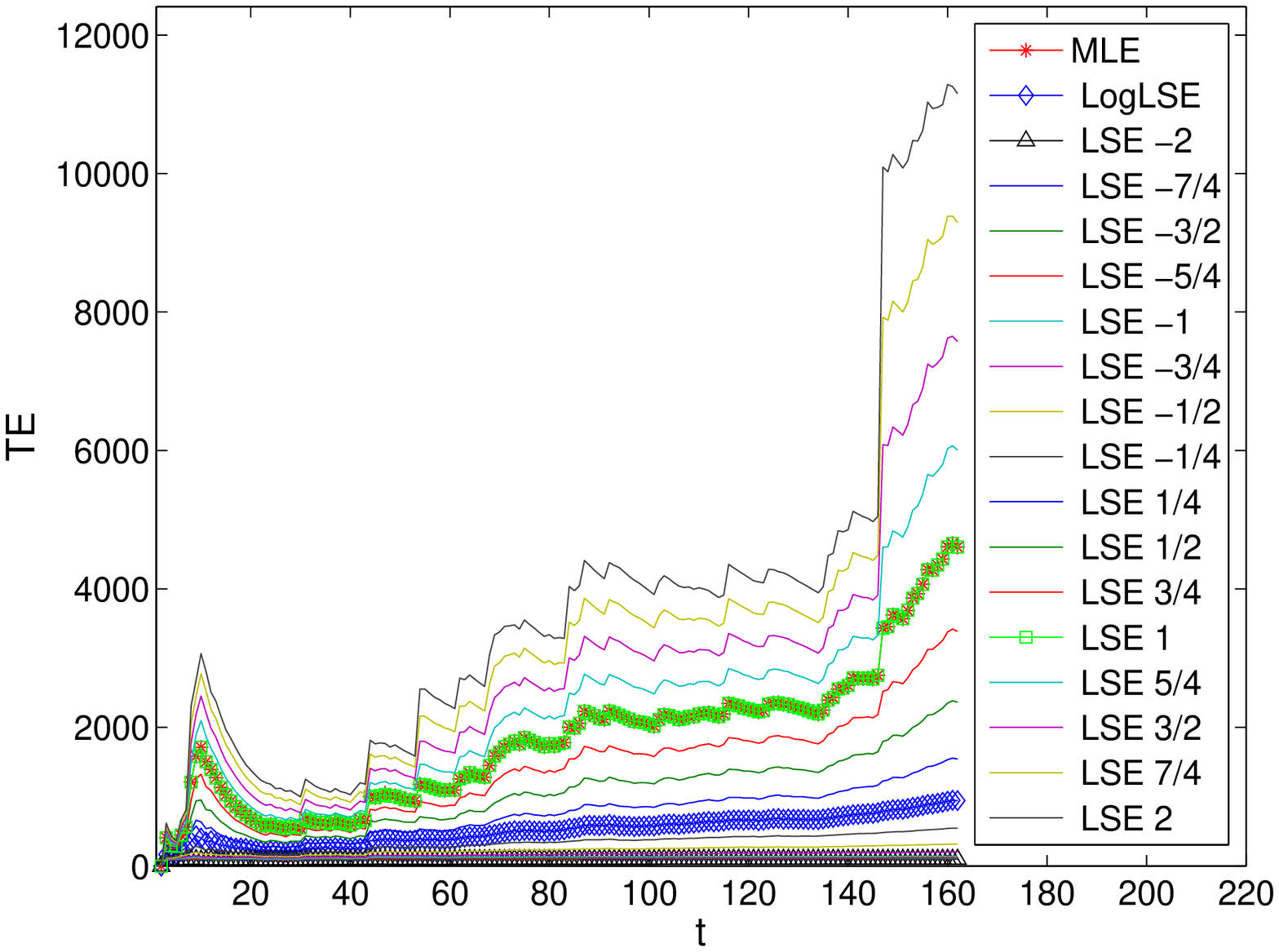}
\caption{TE values of JDM--III with MLE, LogLSE and powLSE.}
\end{minipage}
\end{center}
\end{figure}

\begin{figure}[H]
\begin{center}
\begin{minipage}{5.5cm}
\includegraphics[width=5.5cm]{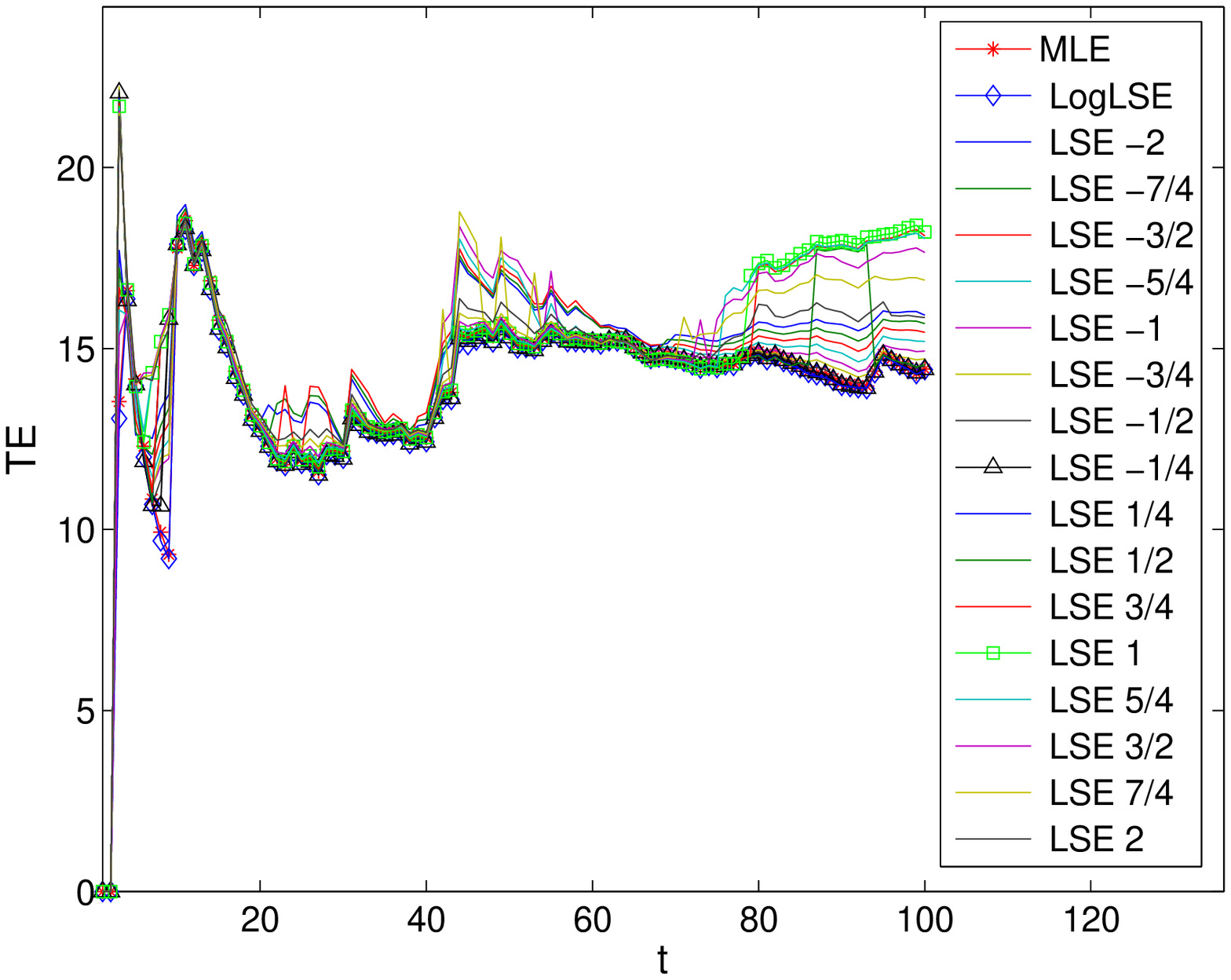}
\caption{TE values of JDM--IV with MLE, LogLSE and powLSE.}
\end{minipage}
\hspace{1cm}
\begin{minipage}{5.5cm}
\includegraphics[width=5.5cm]{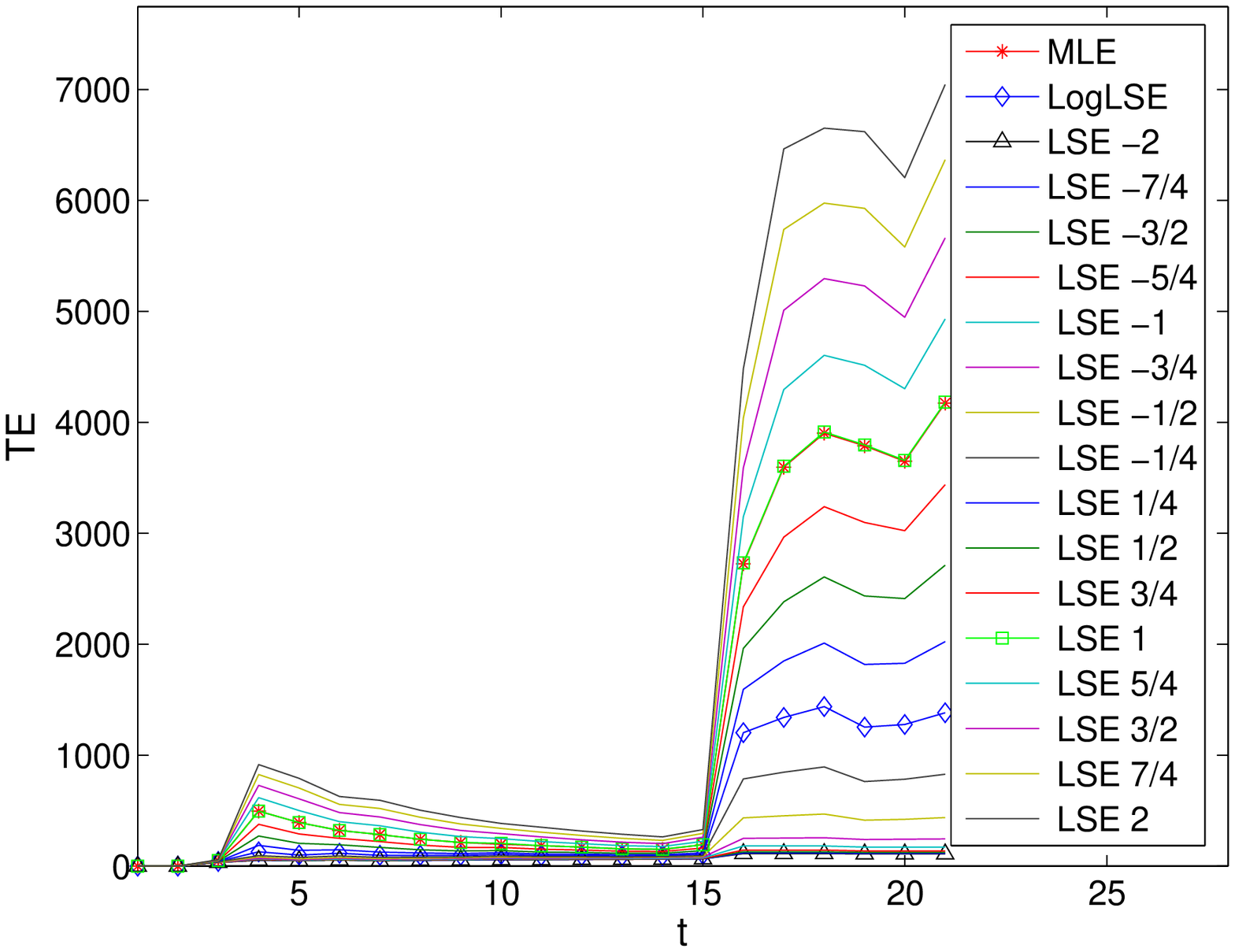}
\caption{TE values of AT\& T with MLE, LogLSE and powLSE.}
\end{minipage}
\end{center}
\end{figure}

From Fig.1 to Fig.6, we can conclude that the TE of powLSE with optimal index along the segmentation data
can achieve relatively small value compared to MLE and LogLSE. From Fig.7 -- Fig.12, we can see that the predictive RE values and the TE values in training data are also similar, hence it demonstrates that  the optimization of power index is reasonable.

\begin{figure}[H]
\begin{center}
\begin{minipage}{5.5cm}
\includegraphics[width=5.5cm]{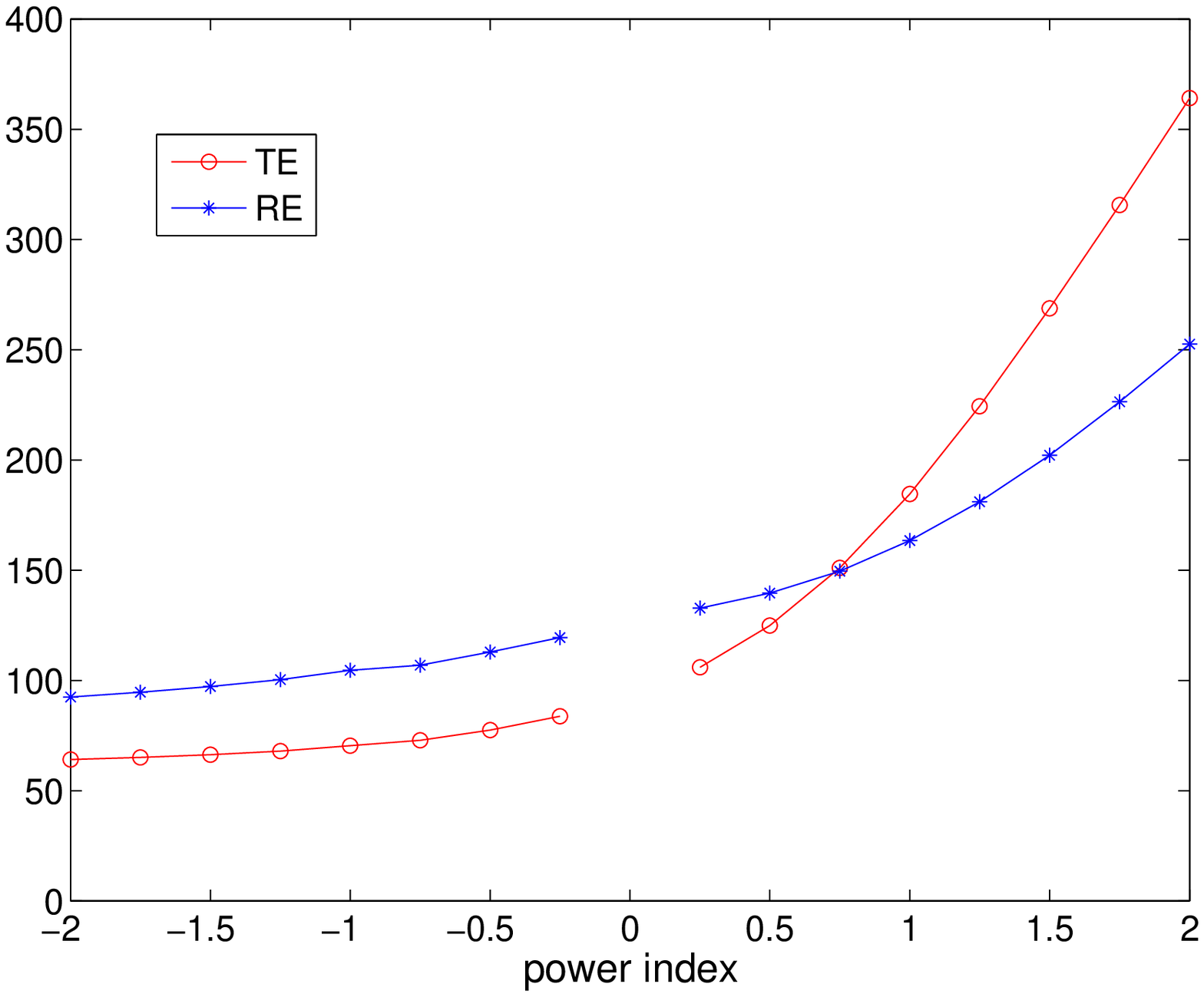}
\caption{RE and TE values of NTDS by powLSE with different power index.}
\end{minipage}
\hspace{1cm}
\begin{minipage}{5.5cm}
\includegraphics[width=5.5cm]{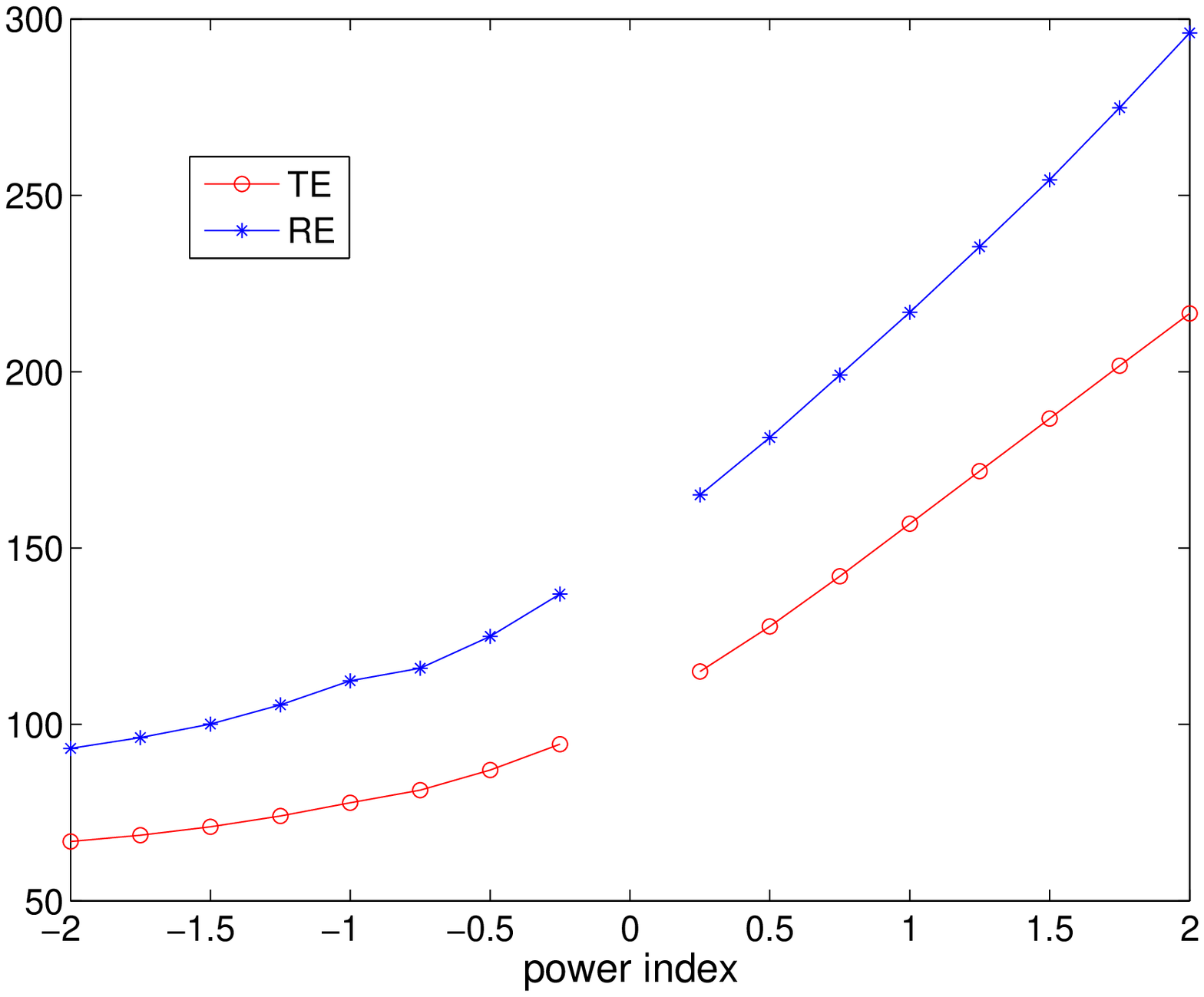}
\caption{RE and TE values of JDM--I by powLSE with different power index.}
\end{minipage}
\end{center}
\end{figure}

\begin{figure}[H]
\begin{center}
\begin{minipage}{5.5cm}
\includegraphics[width=5.5cm]{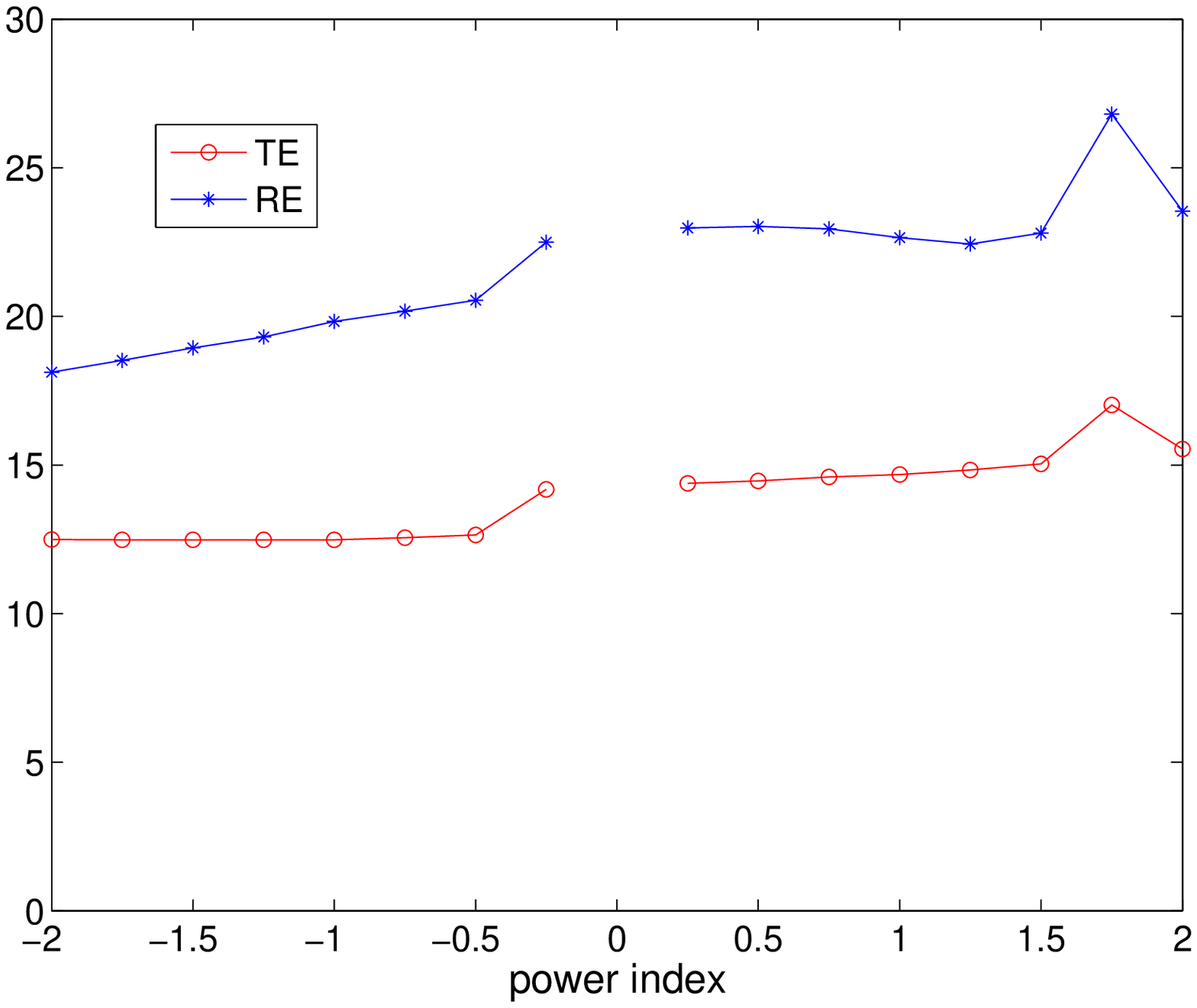}
\caption{RE and TE values of JDM--II by powLSE with different power index.}
\end{minipage}
\hspace{1cm}
\begin{minipage}{5.5cm}
\includegraphics[width=5.5cm]{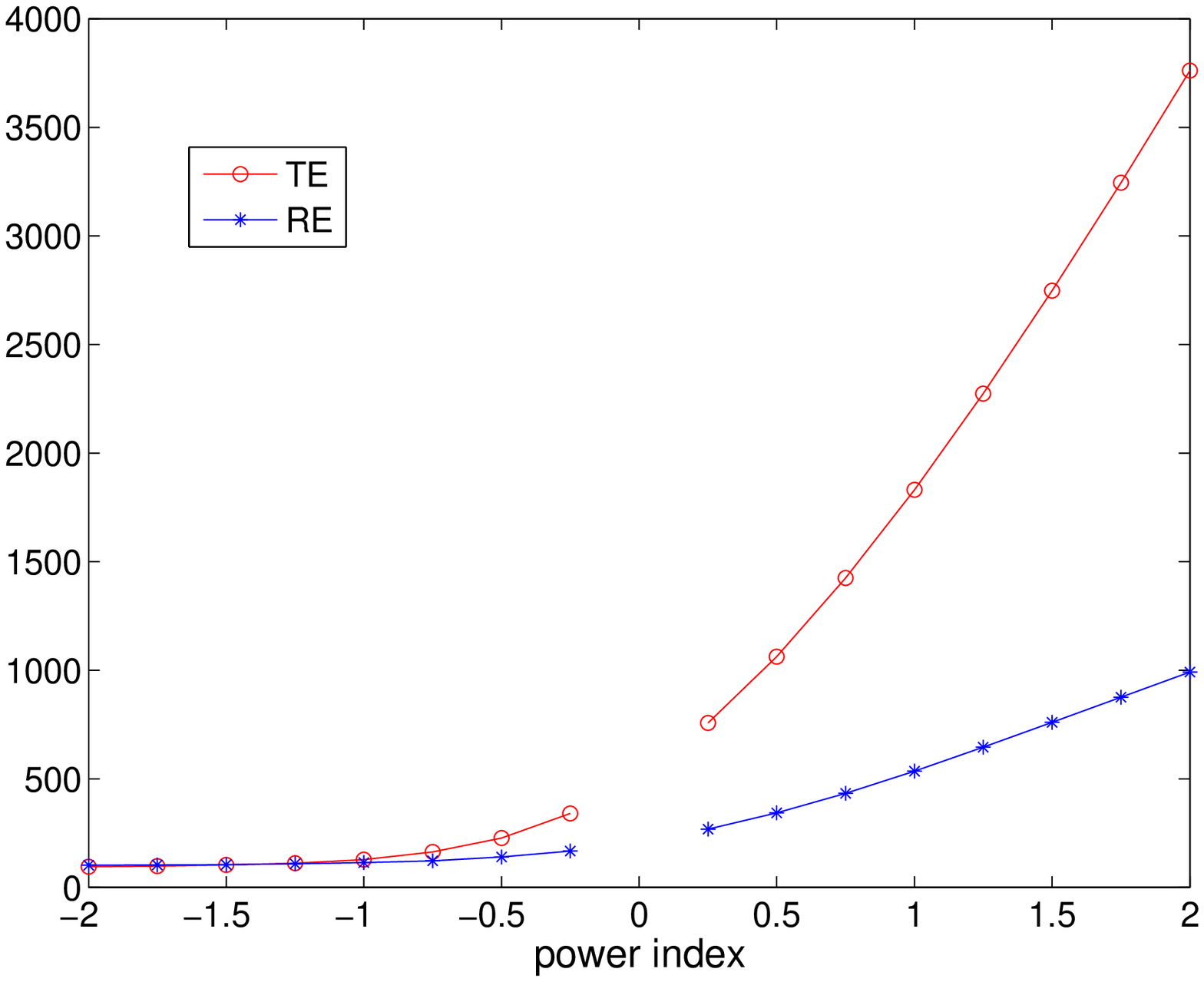}
\caption{RE and TE values of JDM--III by powLSE with different power index.}
\end{minipage}
\end{center}
\end{figure}

\begin{figure}[H]
\begin{center}
\begin{minipage}{5.5cm}
\includegraphics[width=5.5cm]{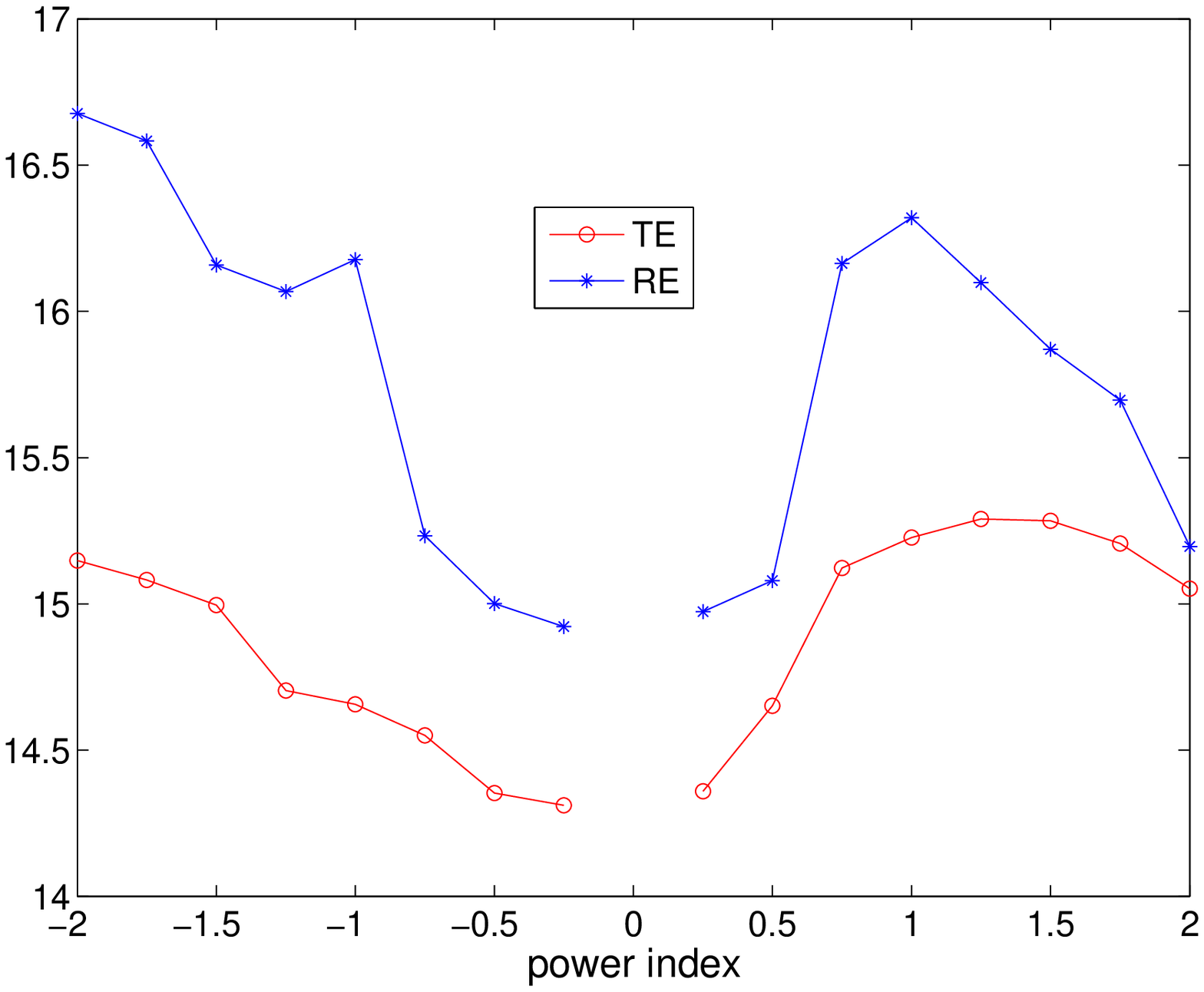}
\caption{RE and TE values of JDM--IV by powLSE with different power index.}
\end{minipage}
\hspace{1cm}
\begin{minipage}{5.5cm}
\includegraphics[width=5.5cm]{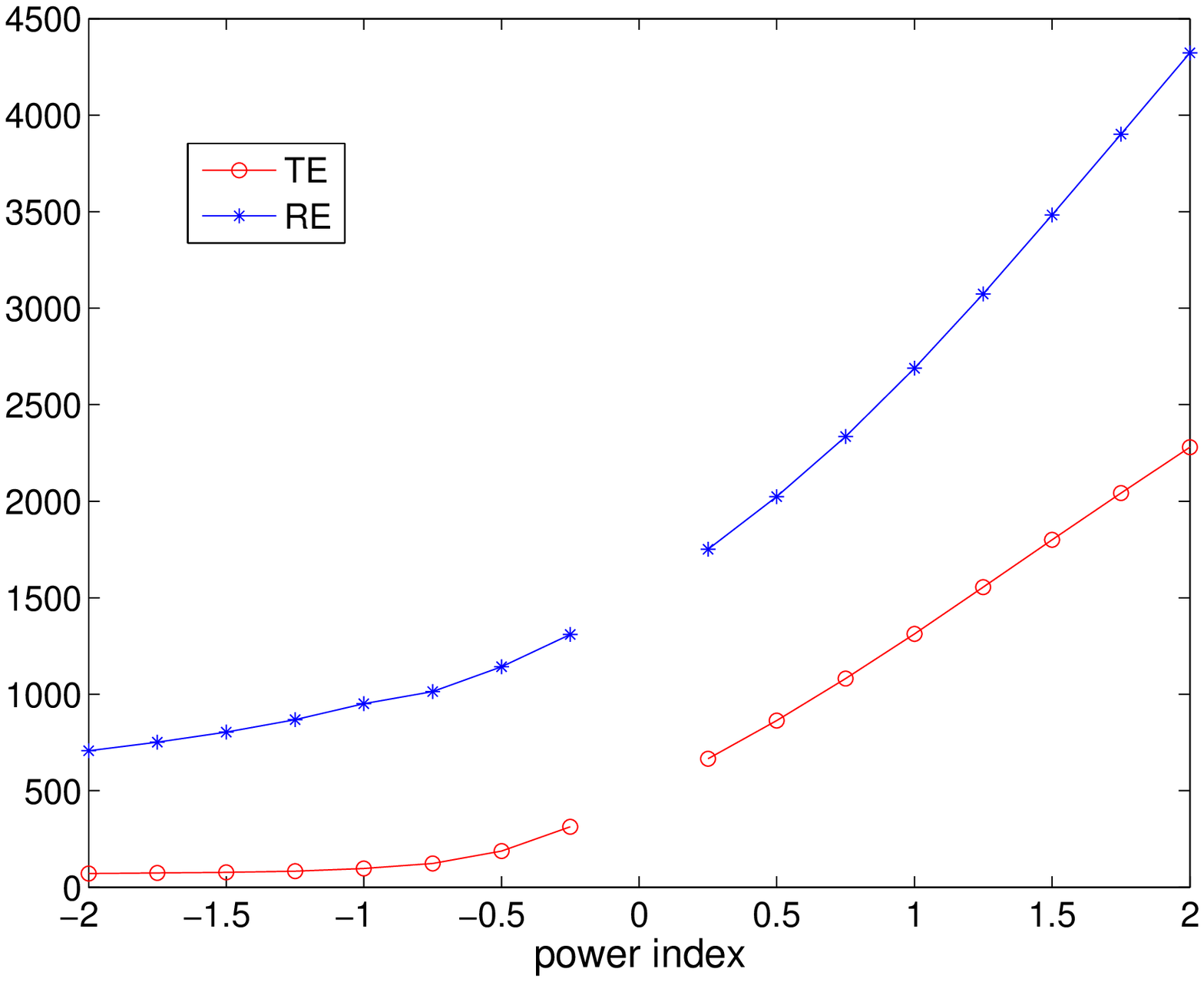}
\caption{RE and TE values of AT\& T by powLSE with different power index.}
\end{minipage}
\end{center}
\end{figure}

\begin{table}[H]
\caption{The RE evaluation of MLE, LSE,LogLSE and powLSE. (\%) }
\label{table:7}
\renewcommand{\tabcolsep}{0.15pc} % enlarge column spacing
\renewcommand{\arraystretch}{0.9} % enlarge line spacing
\begin{center}
\begin{tabular}{lrrrrrr}
\hline
FNLSE  & NTDS & JDM--I & JDM--II & JDM--III & JDM--IV & AT \& T  \\
\hline
MLE            &   162.829 &   216.609 &    21.677 &   536.269 &    16.043 &  2680.787 \\
LSE            &   163.482 &   216.888 &    22.650 &   535.191 &    16.320 &  2688.571 \\
LogLSE         &   125.966 &   150.135 &    21.224 &   208.453 &    16.230 &  1511.177\\
\hline
powLSE opt     &    92.476 &    93.177 &    19.305 &   101.031 &    14.922 &  706.623 \\
$\hat{\alpha}$ &   -2      &-2         &-5/4       & -2        & -1/4      &-2 \\
\hline
powLSE best     &   92.476 &    93.177 &    18.122 &   101.031 &    14.922 &   706.623 \\
$\hat{\alpha}$  &   -2     &    -2     &    -2     &    -2     &    -1/4   &    -2     \\
\hline
\end{tabular}
\end{center}
\end{table}

As powLSE with $\alpha=1 $ is the traditional LSE, the simulation results are
also listed in the Table 7 for comparison.
The RE evaluation results with MLE, LogLSE and powLSE with optimal index in Table 7, the simulation results show
that powLSE with the optimal index can outperform the MLE, LSE and LogLSE according to RE criterion.
And, both  powLSE with optimal index and LogLSE outperform the traditional LSE and MLE.

\subsubsection{Experimental Results with Braun statistic criterion }

The TBS criteria of the recursive training data are shown from Fig. 13  to Fig. 18,
the corresponding one--step--ahead recursive prediction RBS criteria are shown from Fig. 19 -- Fig. 24.
And the RBS values of MLE, LogLSE and powLSE with optimal index are listed in Table 8.

\begin{figure}[H]
\begin{center}
\begin{minipage}{5.5cm}
\includegraphics[width=5.5cm]{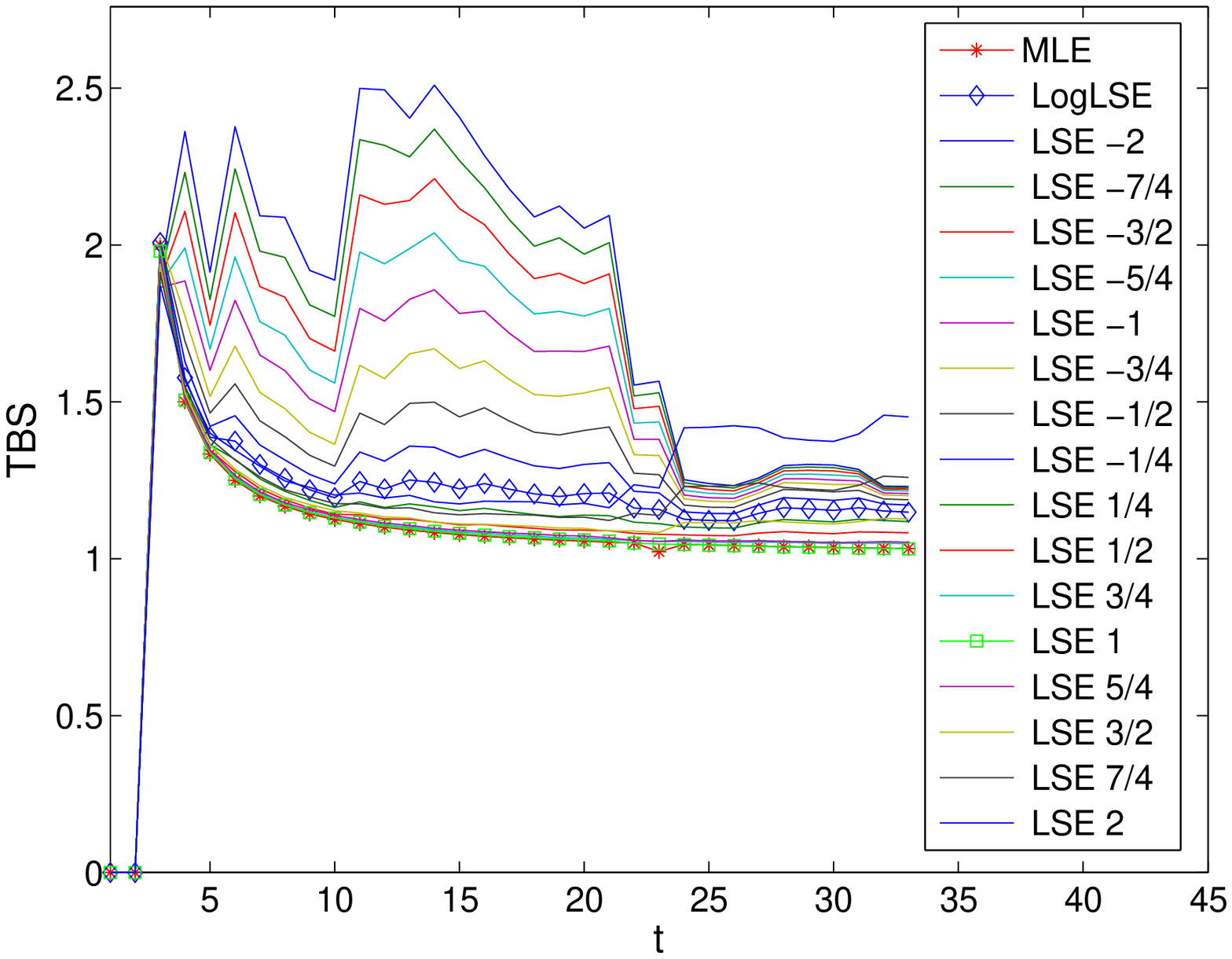}
\caption{TBR values of NTDS with MLE, LogLSE and powLSE.}
\end{minipage}
\hspace{1cm}
\begin{minipage}{5.5cm}
\includegraphics[width=5.5cm]{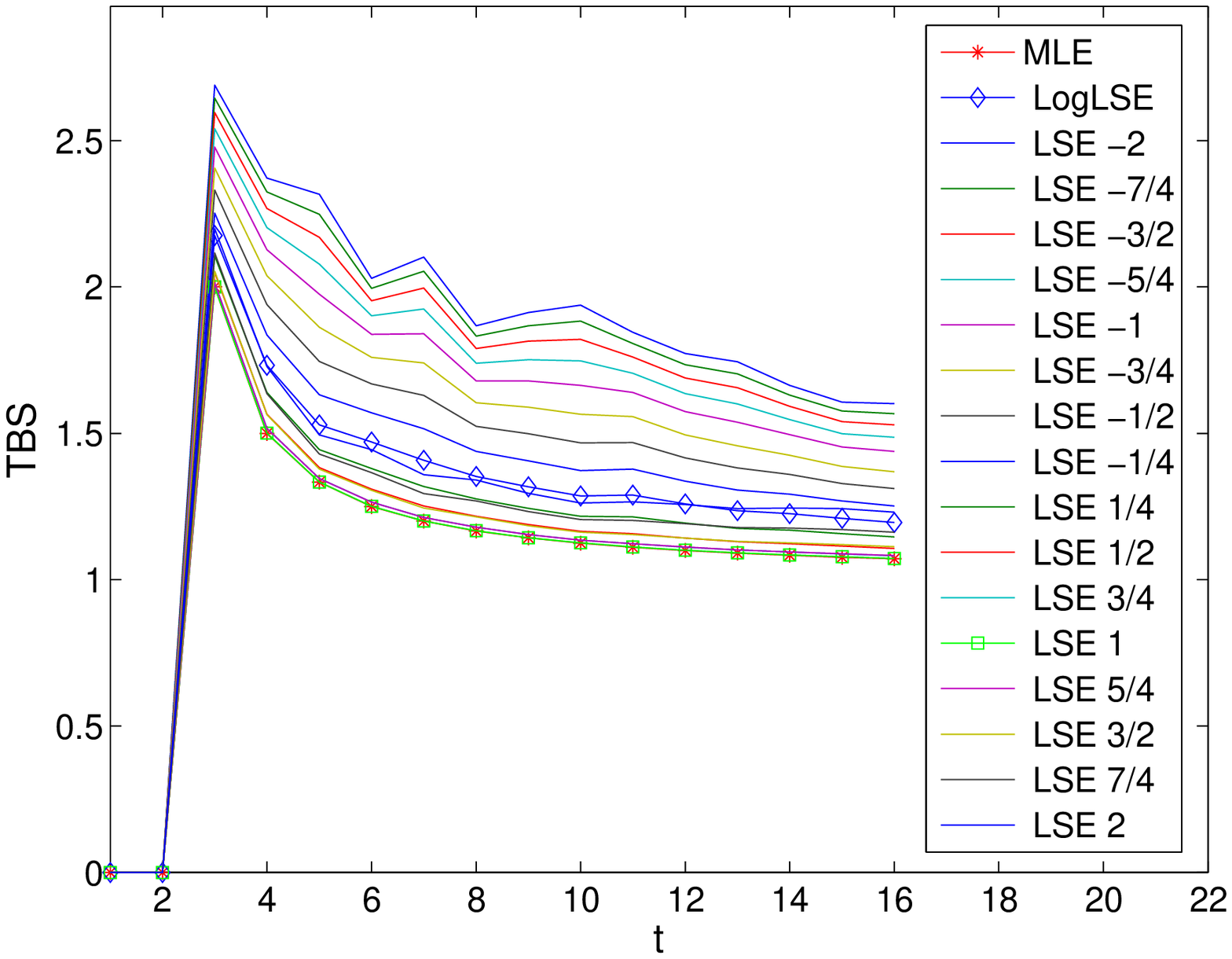}
\caption{TBR values of JDM--I with MLE, LogLSE and powLSE.}
\end{minipage}
\end{center}
\end{figure}

\begin{figure}[H]
\begin{center}
\begin{minipage}{5.5cm}
\includegraphics[width=5.5cm]{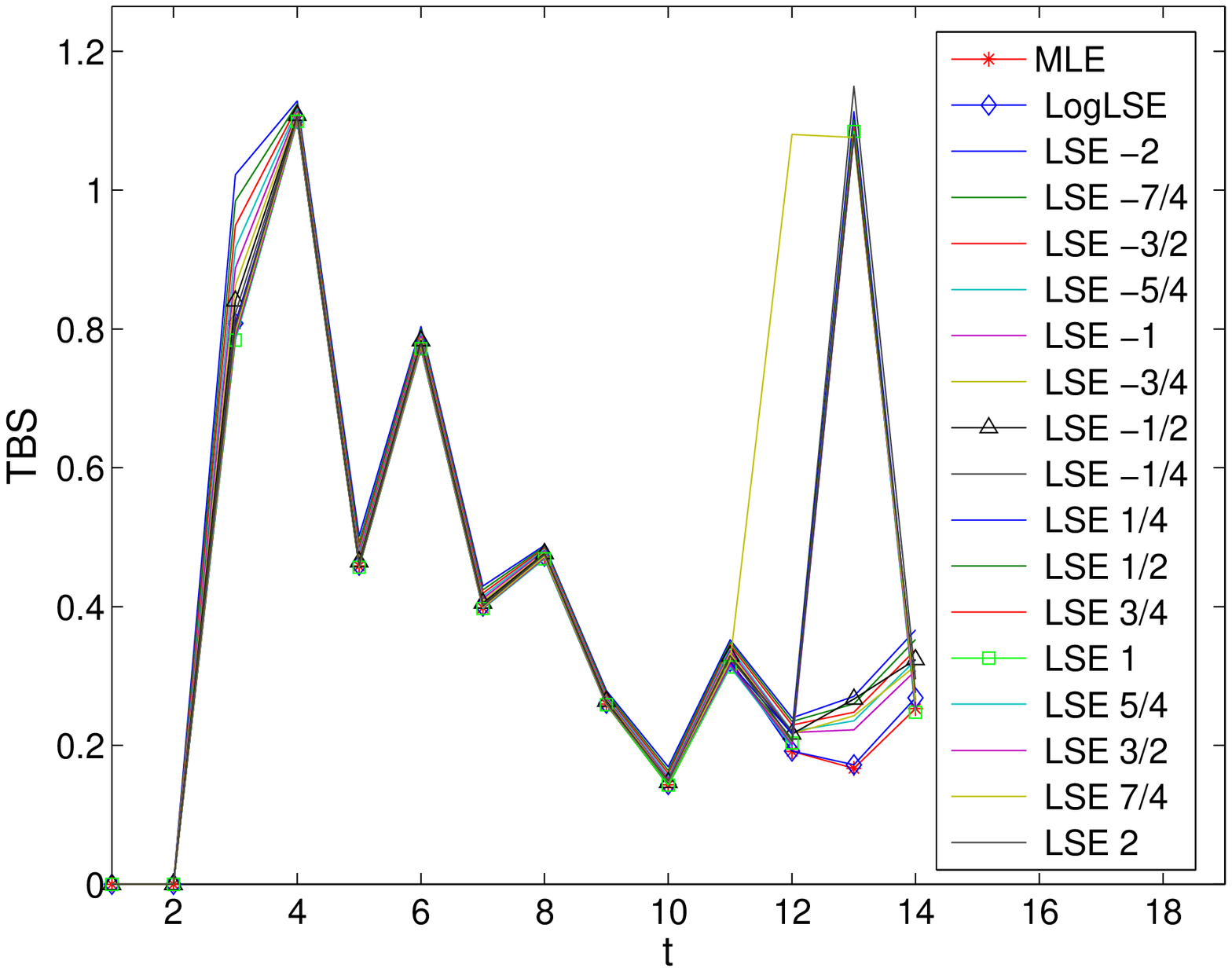}
\caption{TBR values of JDM--II with MLE, LogLSE and powLSE.}
\end{minipage}
\hspace{1cm}
\begin{minipage}{5.5cm}
\includegraphics[width=5.5cm]{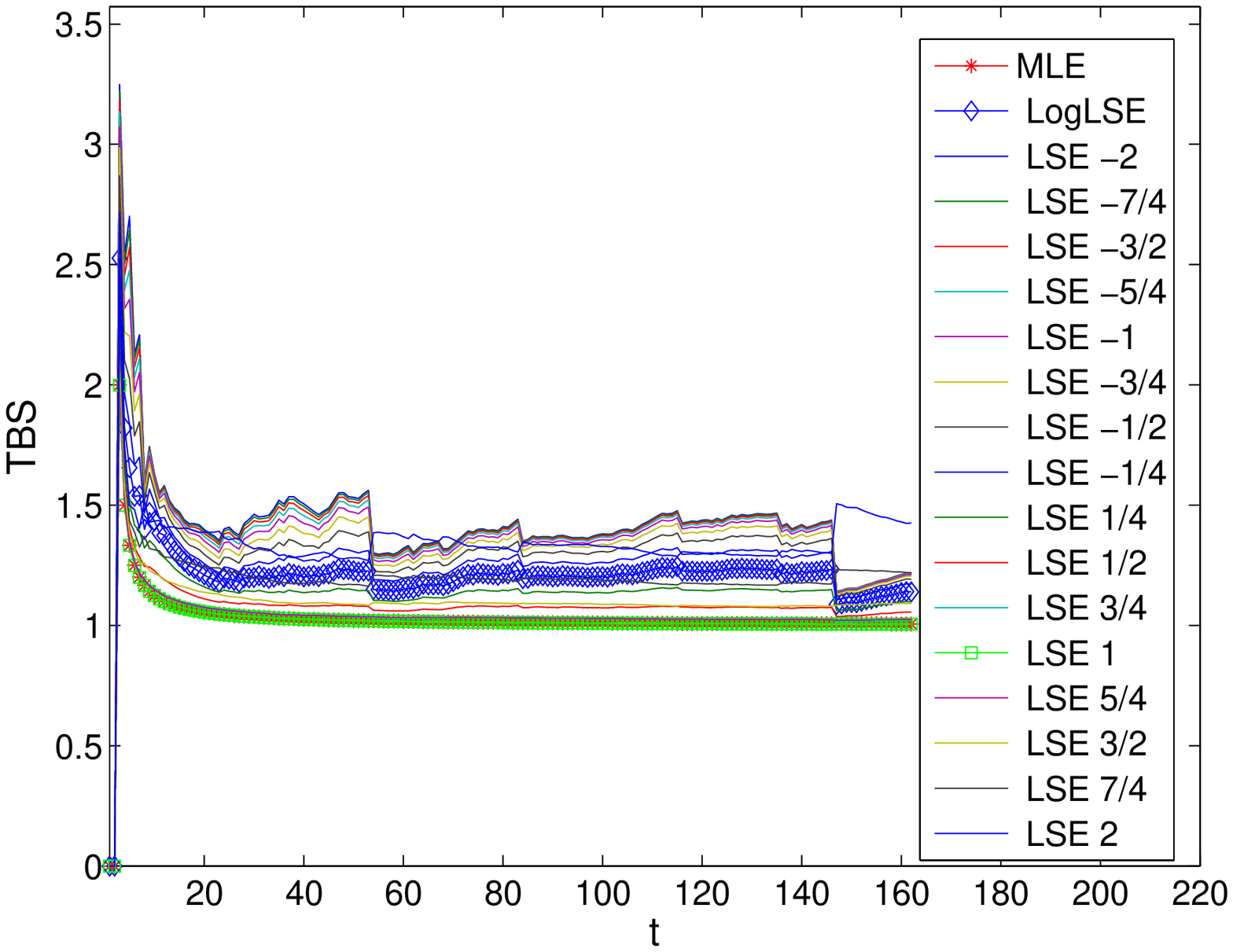}
\caption{TBR values of JDM--III with MLE, LogLSE and powLSE.}
\end{minipage}
\end{center}
\end{figure}

\begin{figure}[H]
\begin{center}
\begin{minipage}{5.5cm}
\includegraphics[width=5.5cm]{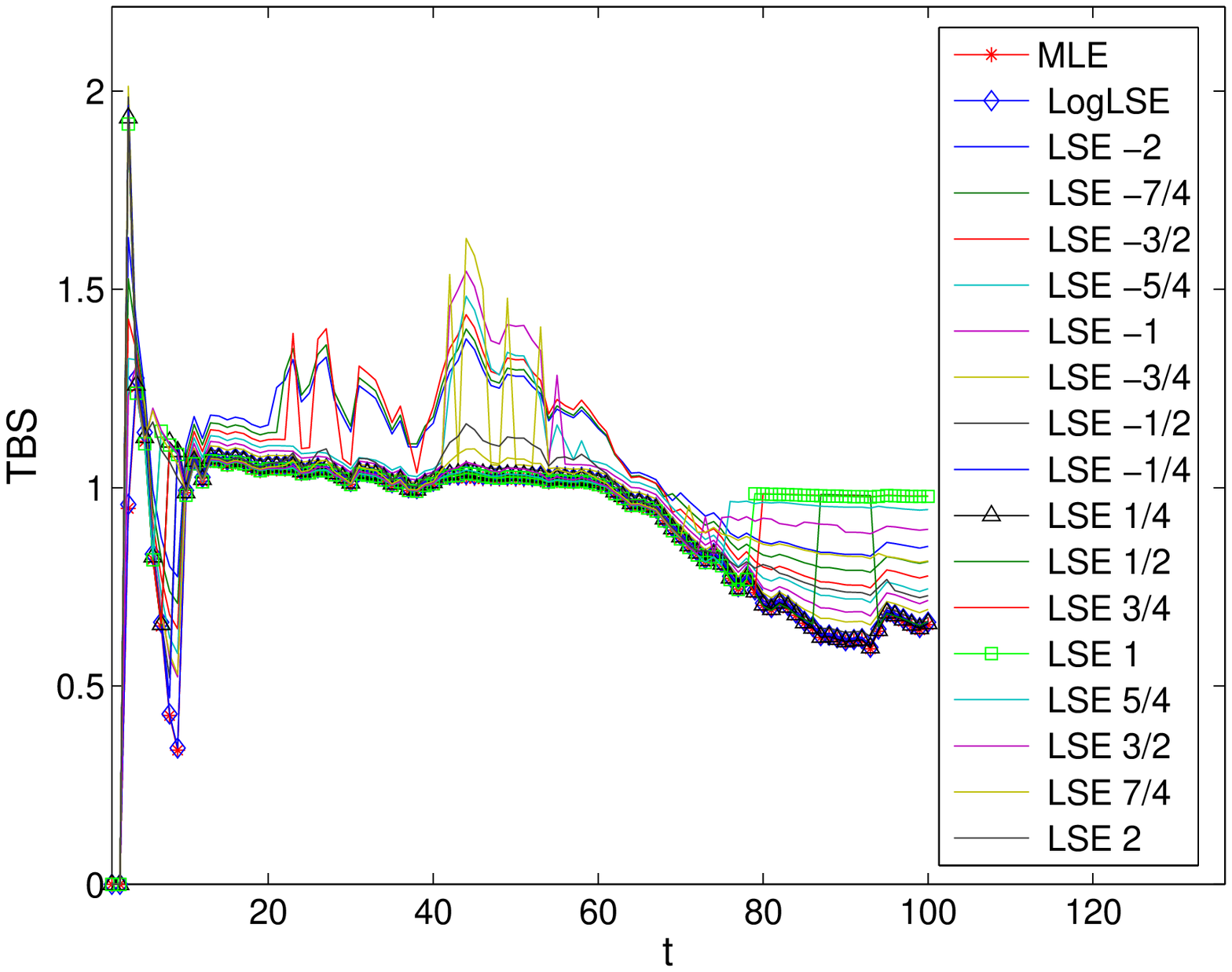}
\caption{TBR values of JDM--IV with MLE, LogLSE and powLSE.}
\end{minipage}
\hspace{1cm}
\begin{minipage}{5.5cm}
\includegraphics[width=5.5cm]{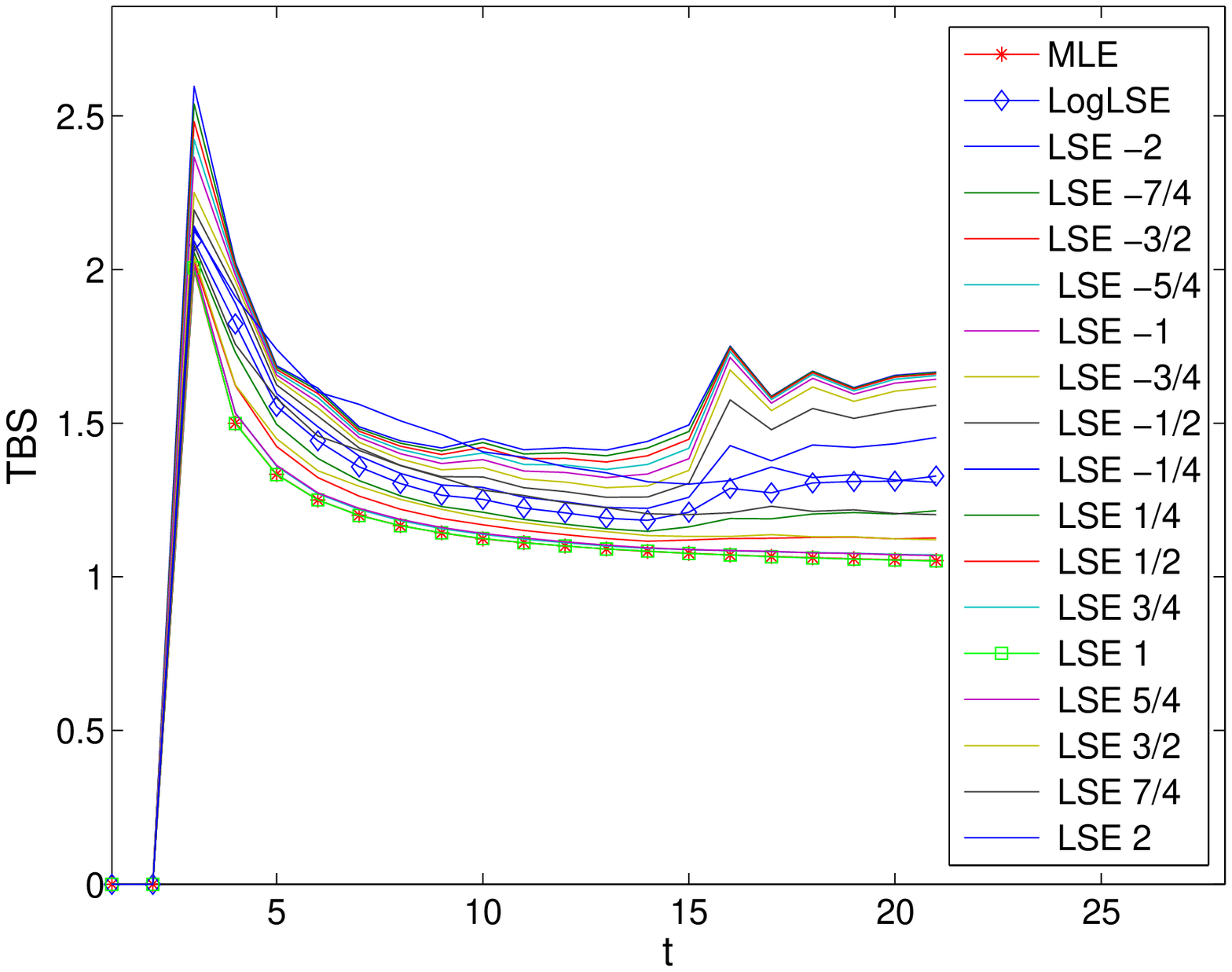}
\caption{TBR values of AT\& T with MLE, LogLSE and powLSE.}
\end{minipage}
\end{center}
\end{figure}

From Fig. 13  to Fig. 18, we can conclude that powLSE with optimal index can achieve relatively smaller profiles than
MLE, LSE ad LogLSE. And, Fig. 19 --Fig. 24 manifest the validation of power index optimization with Braun statistic criterion.

\begin{figure}[H]
\begin{center}
\begin{minipage}{5.5cm}
\includegraphics[width=5.5cm]{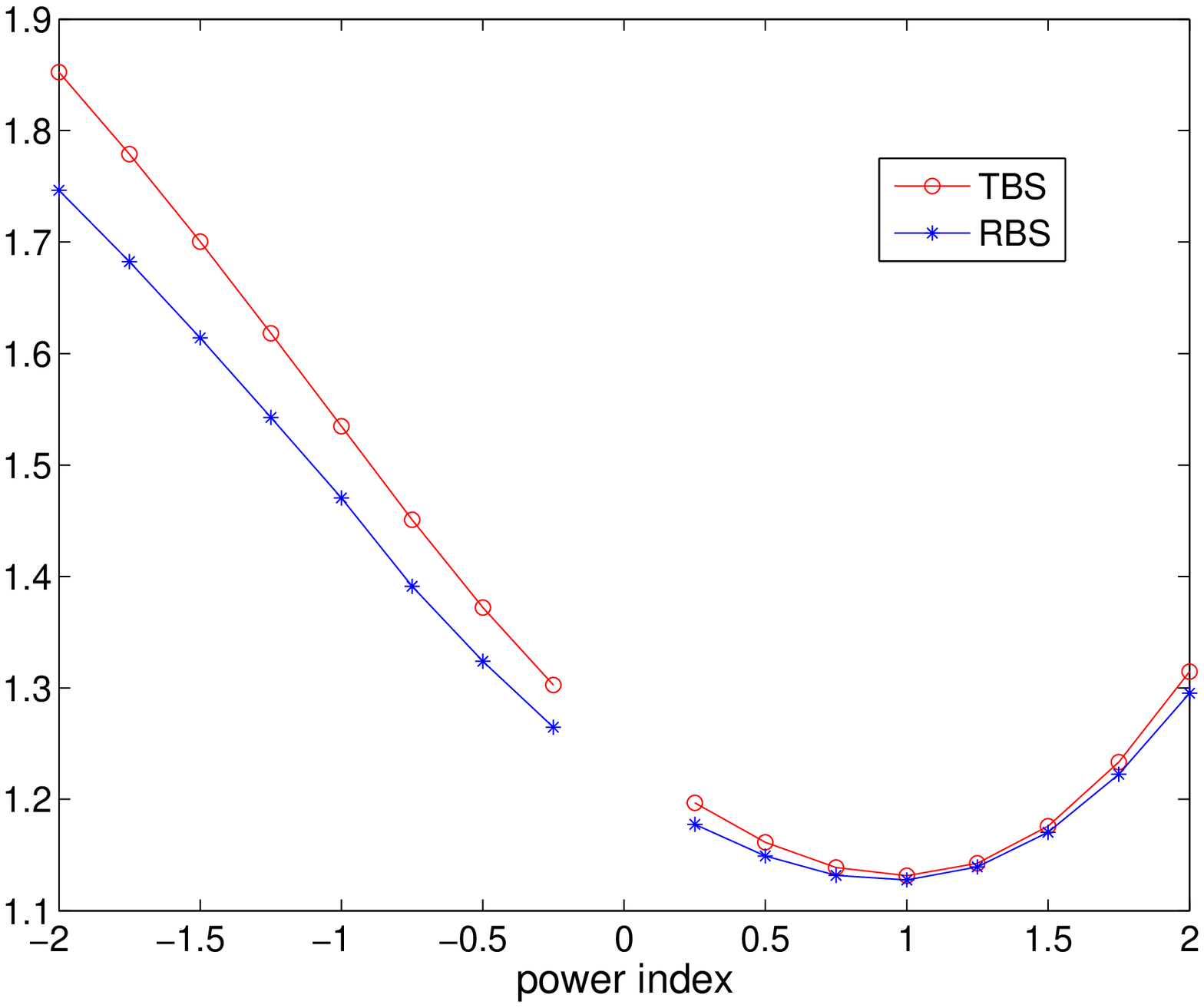}
\caption{RBS and TBS values of NTDS by powLSE with different power index.}
\end{minipage}
\hspace{1cm}
\begin{minipage}{5.5cm}
\includegraphics[width=5.5cm]{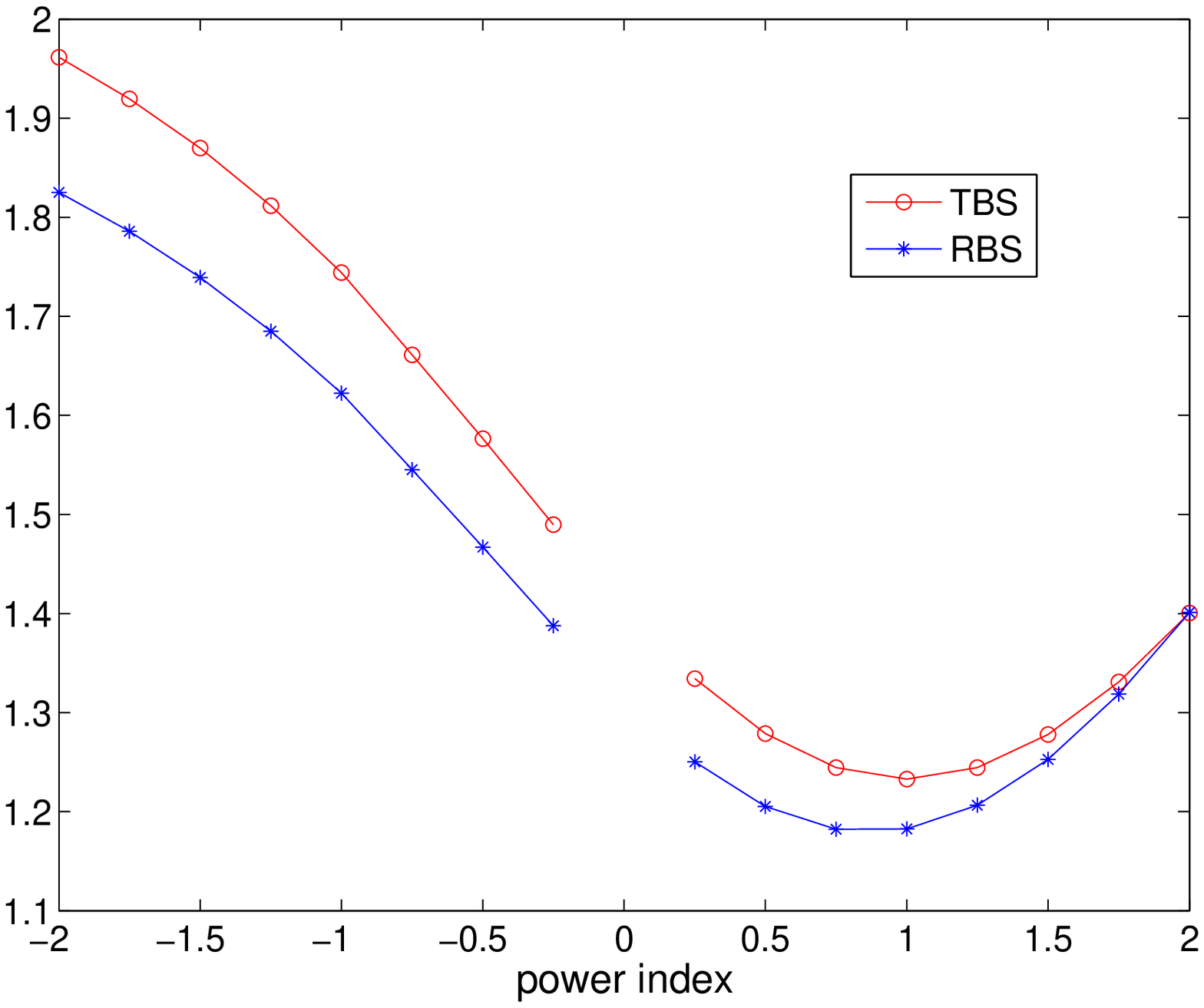}
\caption{RBS and TBS values of JDM--I by powLSE with different power index.}
\end{minipage}
\end{center}
\end{figure}

\begin{figure}[H]
\begin{center}
\begin{minipage}{5.5cm}
\includegraphics[width=5.5cm]{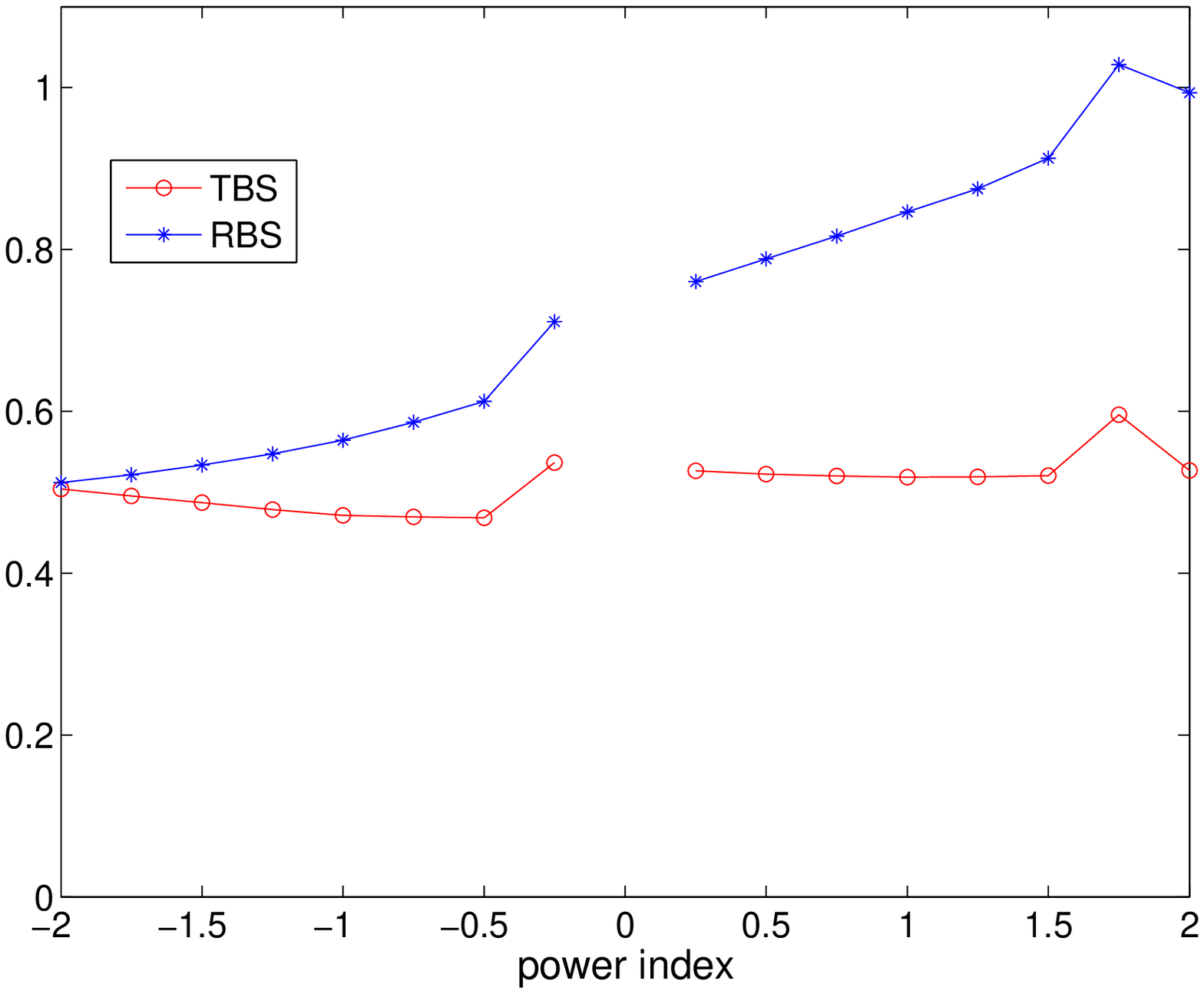}
\caption{RBS and TBS values of JDM--II by powLSE with different power index.}
\end{minipage}
\hspace{1cm}
\begin{minipage}{5.5cm}
\includegraphics[width=5.5cm]{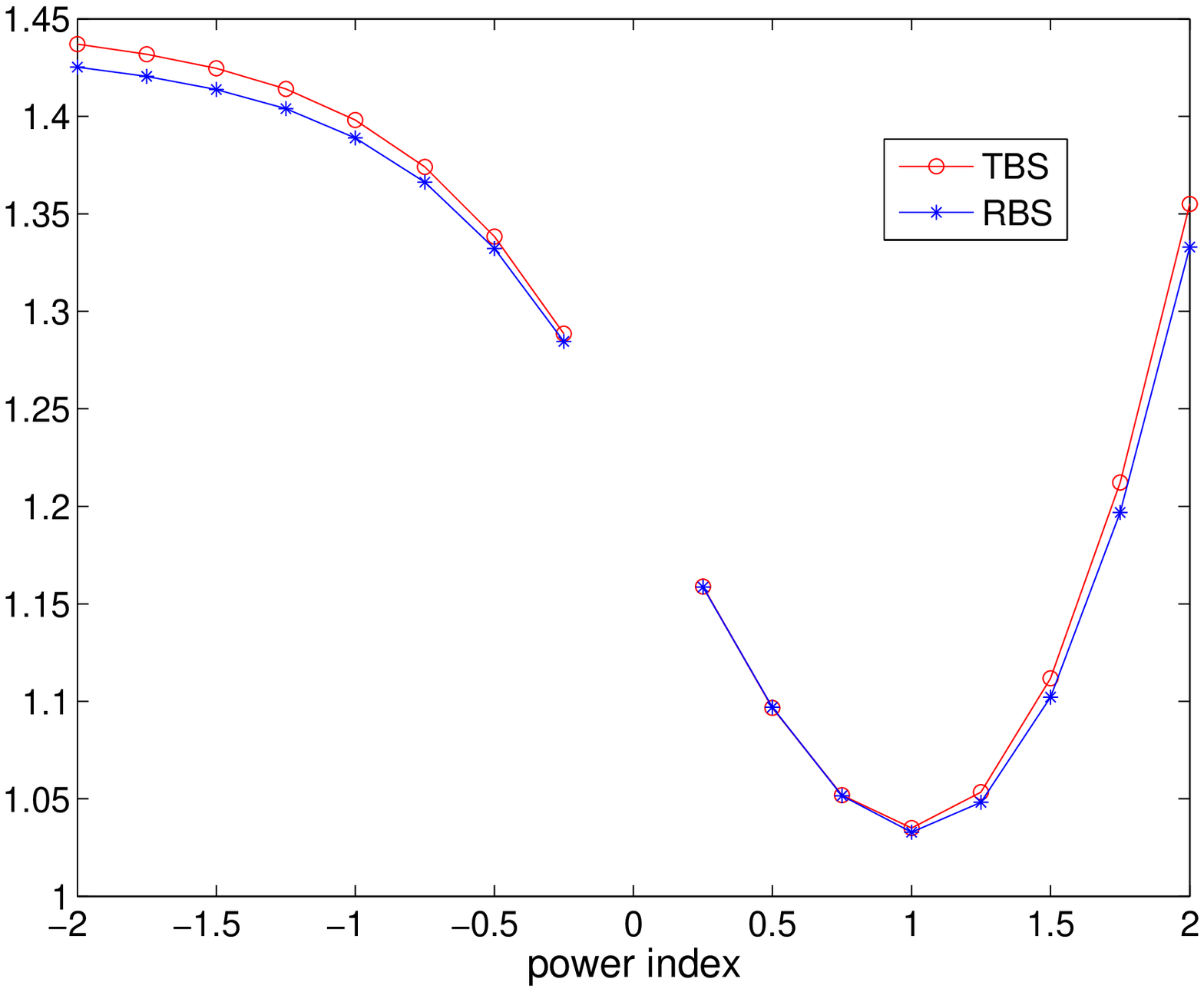}
\caption{RBS and TBS values of JDM--III by powLSE with different power index.}
\end{minipage}
\end{center}
\end{figure}

\begin{figure}[H]
\begin{center}
\begin{minipage}{5.5cm}
\includegraphics[width=5.5cm]{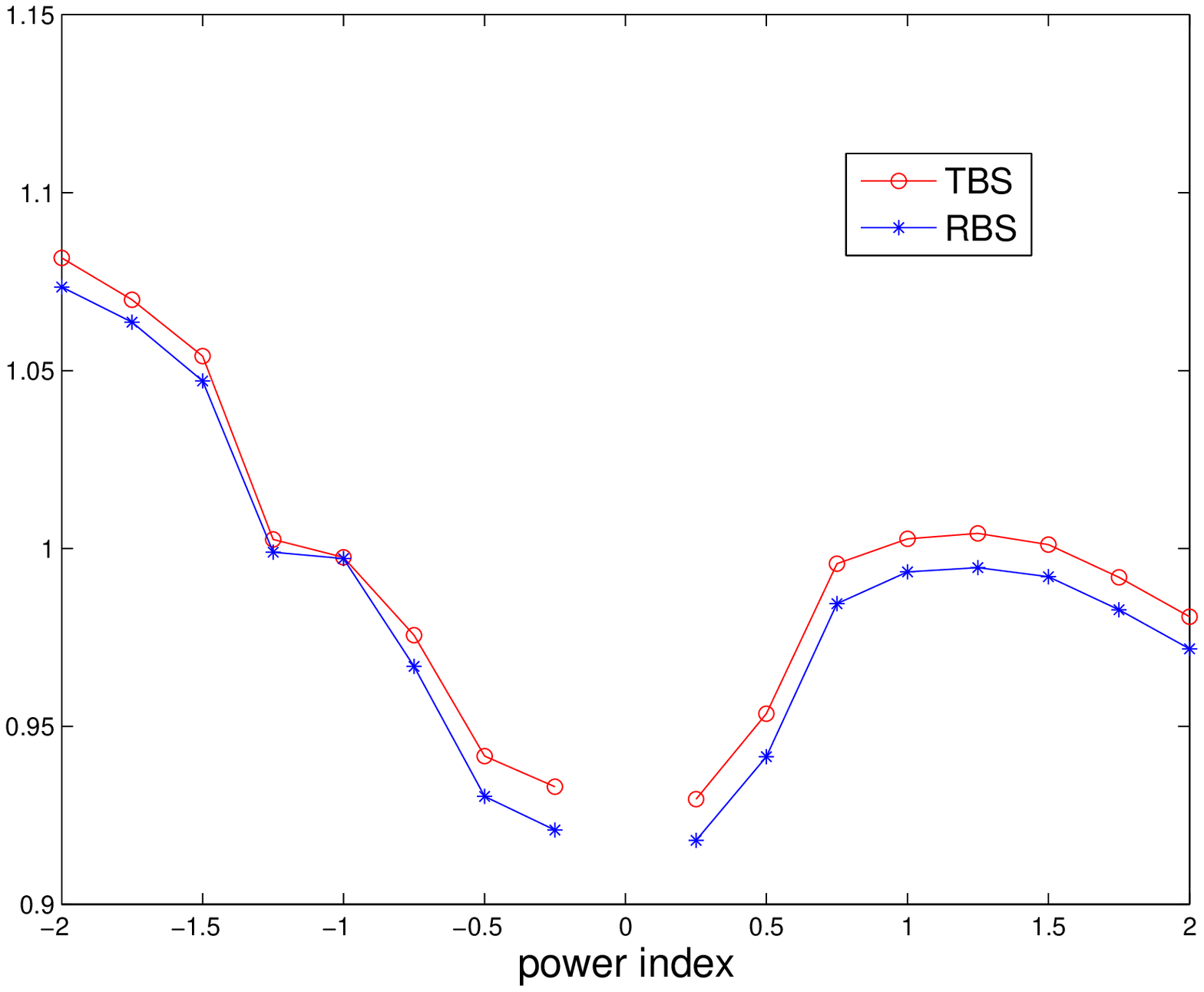}
\caption{RBS and TBS values of JDM--IV by powLSE with different power index.}
\end{minipage}
\hspace{1cm}
\begin{minipage}{5.5cm}
\includegraphics[width=5.5cm]{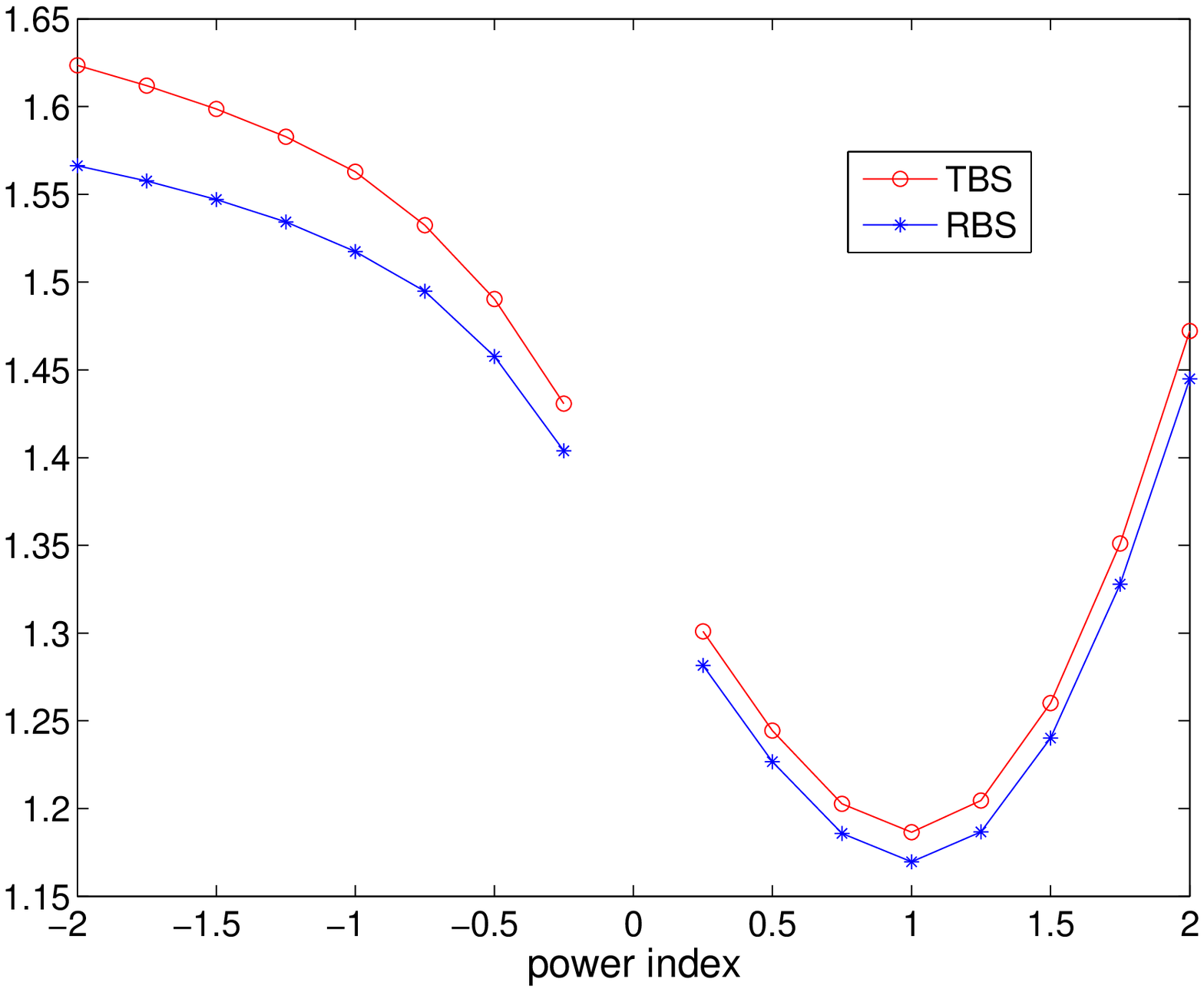}
\caption{RBS and TBS values of AT\& T by powLSE with different power index.}
\end{minipage}
\end{center}
\end{figure}

\begin{table}[H]
\caption{The Braun statistic evaluation of LSE,LogLSE and powLSE. (\%) }
\label{table:8}
\renewcommand{\tabcolsep}{0.15pc} % enlarge column spacing
\renewcommand{\arraystretch}{0.9} % enlarge line spacing
\begin{center}
\begin{tabular}{lrrrrrr}
\hline
FNLSE  & NTDS & JDM--I & JDM--II & JDM--III & JDM--IV & AT \& T  \\
\hline
MLE &     1.124 &     1.182 &     0.657 &     1.033 &     0.955 &     1.170 \\
LSE &     1.128 &     1.183 &     0.847 &     1.033 &     0.994 &     1.170 \\
LogLSE &     1.216 &     1.313 &     0.653 &     1.225 &     0.963 &     1.342 \\
\hline
powLSE opt     &     1.128 &     1.183 &     0.612 &     1.033 &     0.918 &     1.170 \\
$\hat{\alpha}$ &     1 &     1 &    -1/2 &     1 &     1/4 &     1 \\
\hline
powLSE best     &     1.128 &     1.182 &     0.512 &     1.033 &     0.918 &     1.170 \\
$\hat{\alpha}$  &     1 &     3/4 &    -2 &     1 &     1/4 &     1 \\
\hline
\end{tabular}
\end{center}
\end{table}

From the RBS evaluation values in Table 8, we can see that powLSE with optimal index outperforms LSE on data set JDM--II and  JDM--IV, and has same performances comparing with LSE on the other four data sets. It demonstrates that
powLSE  extends LSE according to Braun statistic criterion.

\subsubsection{Heteroscedasticity of data}

All of the corresponding variances with original data and residual data predicted by MLE, LSE ,LogLSE and powLSE with optimal index are shown in Fig. 25 -- Fig. 30, where we denote the powLSE optimized by RE criterion as powLSE RE, and denote the powLSE optimized by Braun statistic criterion as powLSE BS.

\begin{figure}[H]
\begin{center}
\begin{minipage}{5.5cm}
\includegraphics[width=5.5cm]{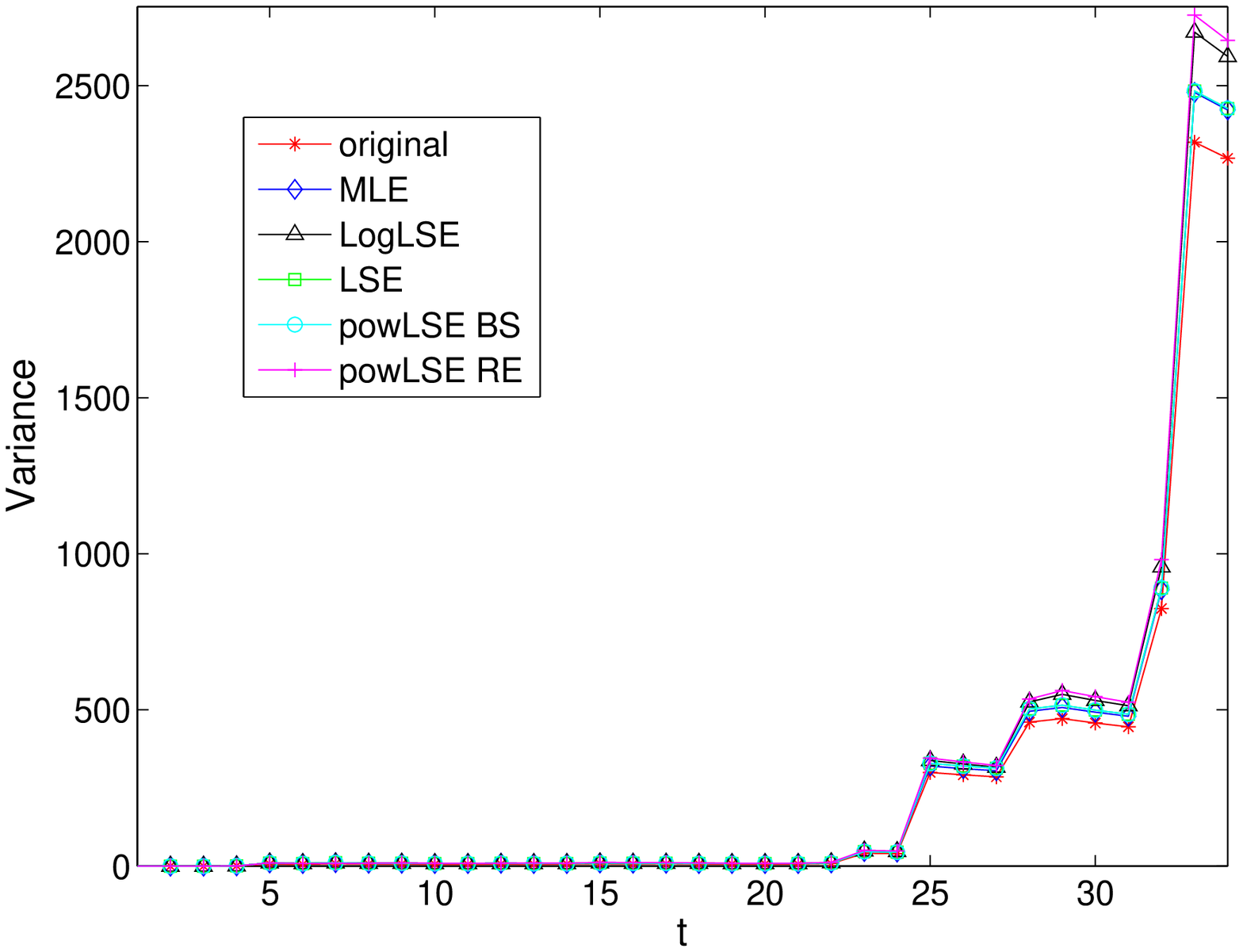}
\caption{Variances of NTDS with MLE, LogLSE and powLSE.}
\end{minipage}
\hspace{1cm}
\begin{minipage}{5.5cm}
\includegraphics[width=5.5cm]{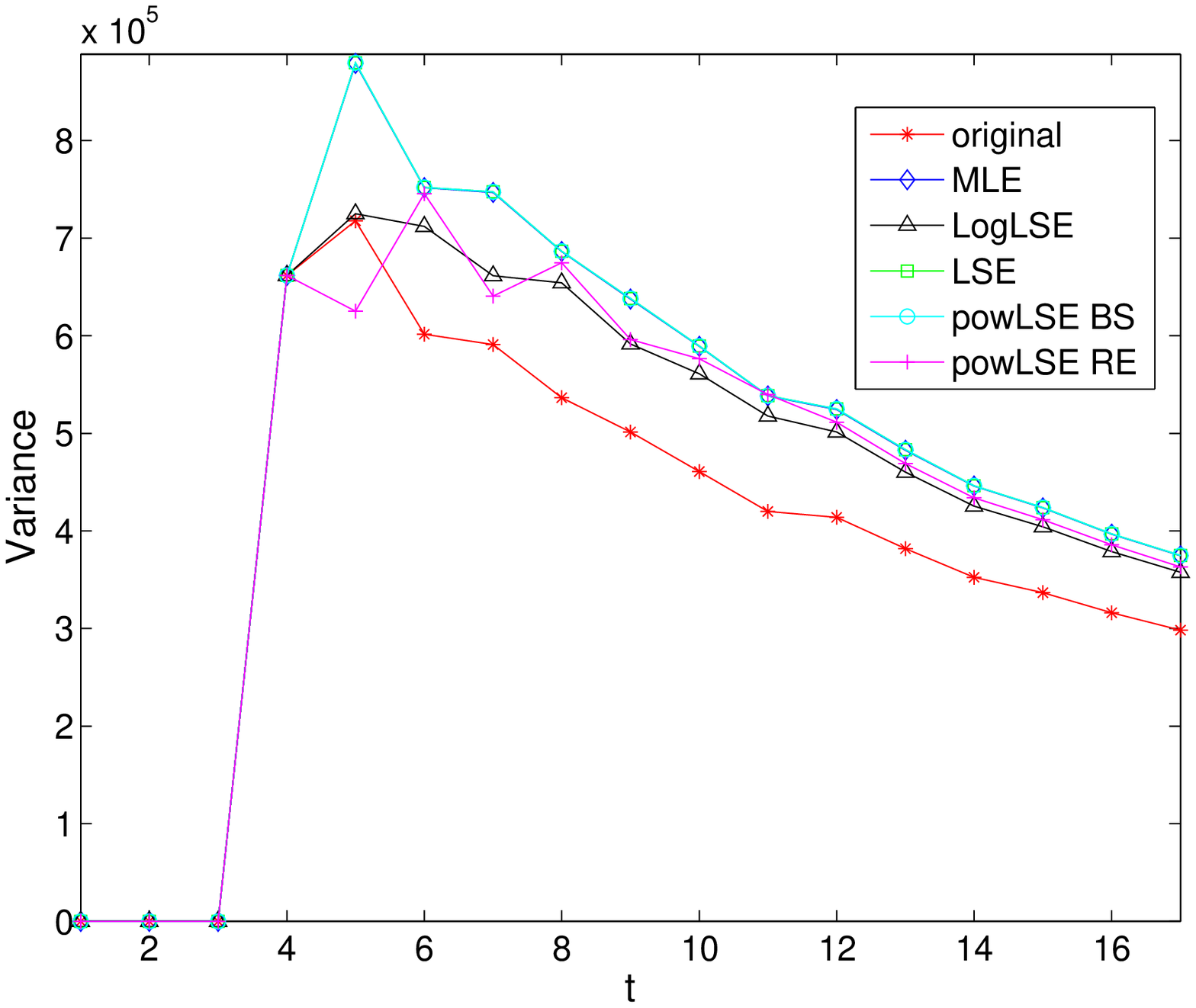}
\caption{Variances of JDM--I with MLE, LogLSE and powLSE.}
\end{minipage}
\end{center}
\end{figure}

\begin{figure}[H]
\begin{center}
\begin{minipage}{5.5cm}
\includegraphics[width=5.5cm]{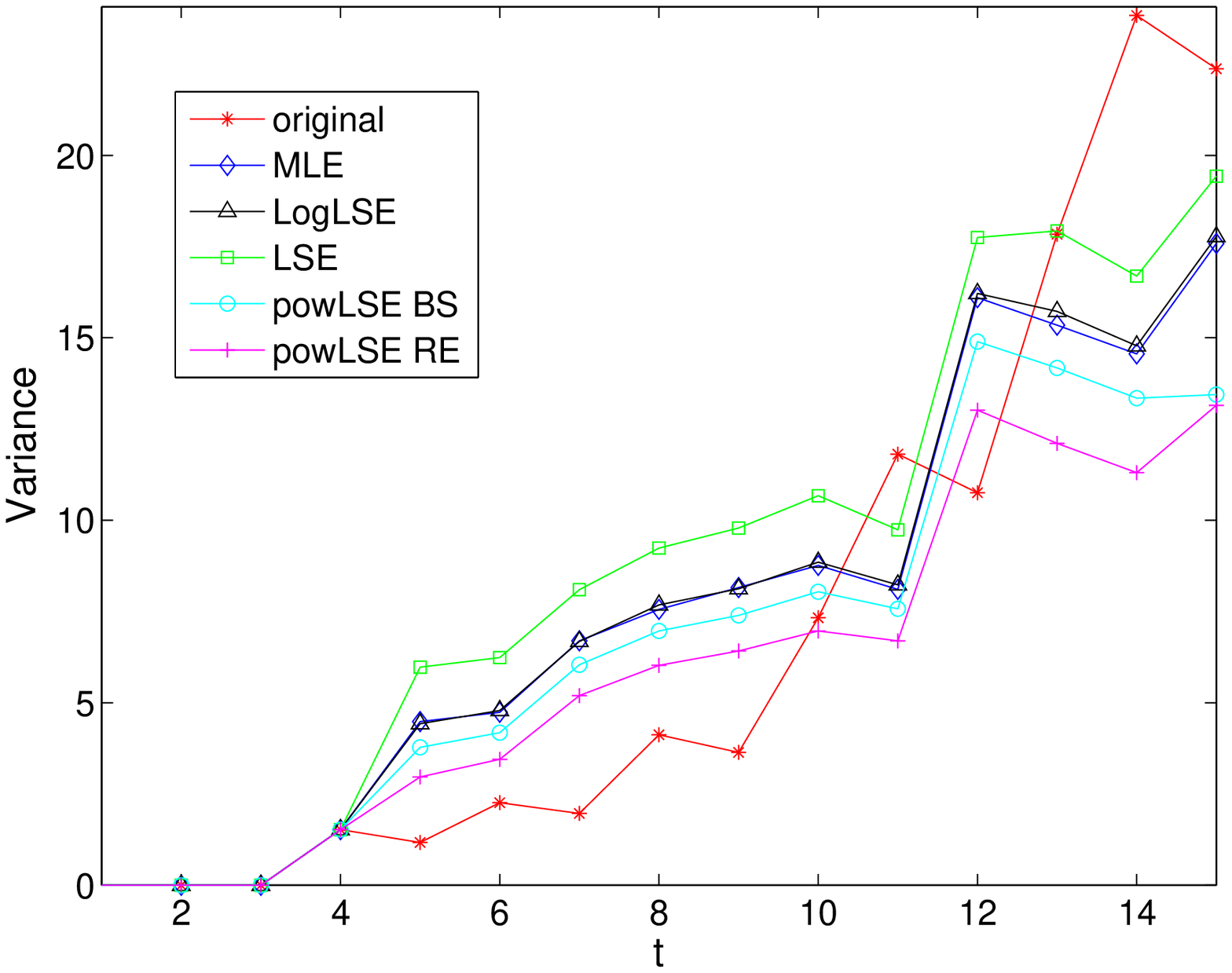}
\caption{Variances of JDM--II with MLE, LogLSE and powLSE.}
\end{minipage}
\hspace{1cm}
\begin{minipage}{5.5cm}
\includegraphics[width=5.5cm]{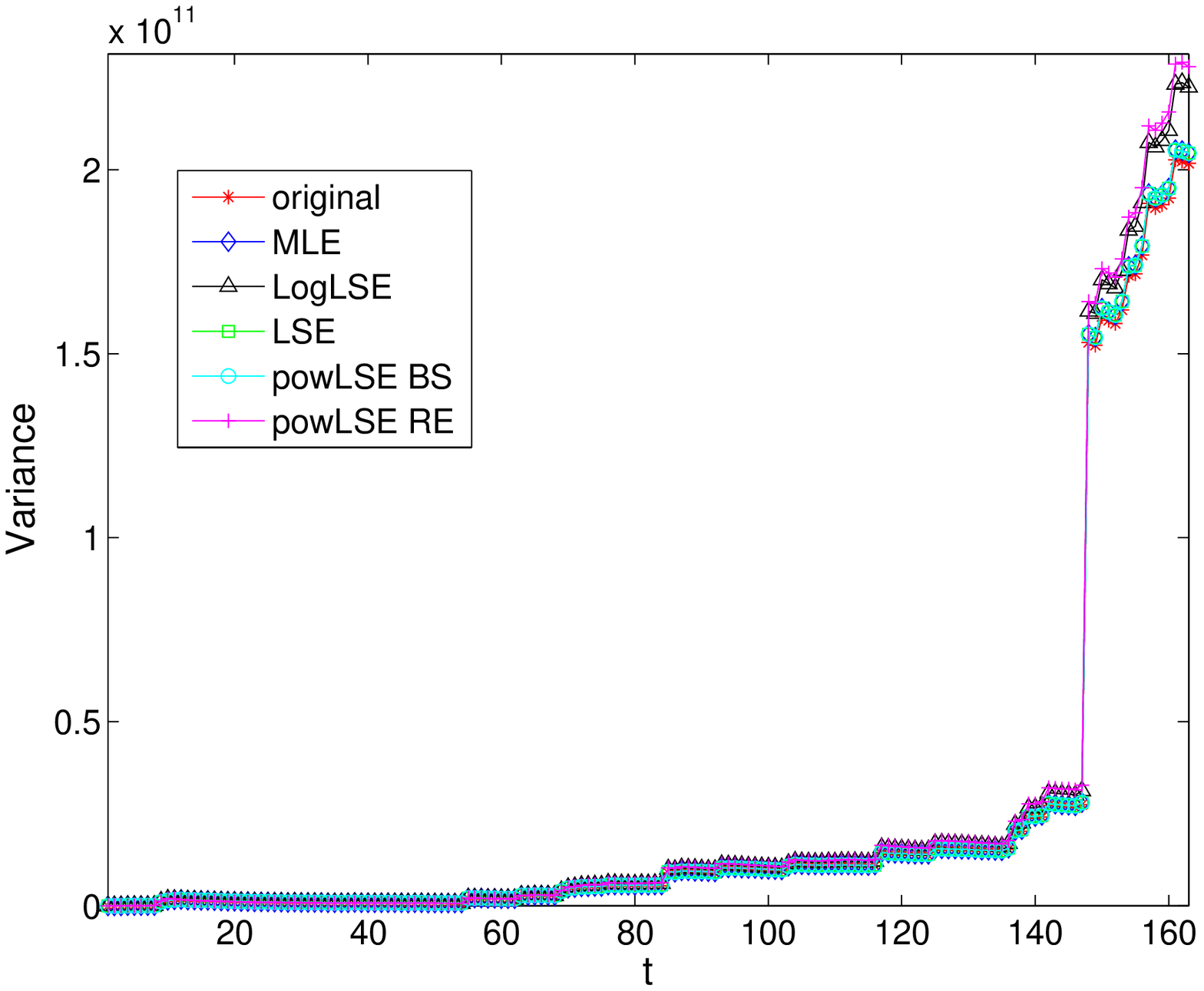}
\caption{Variances of JDM--III with MLE, LogLSE and powLSE.}
\end{minipage}
\end{center}
\end{figure}

\begin{figure}[H]
\begin{center}
\begin{minipage}{5.5cm}
\includegraphics[width=5.5cm]{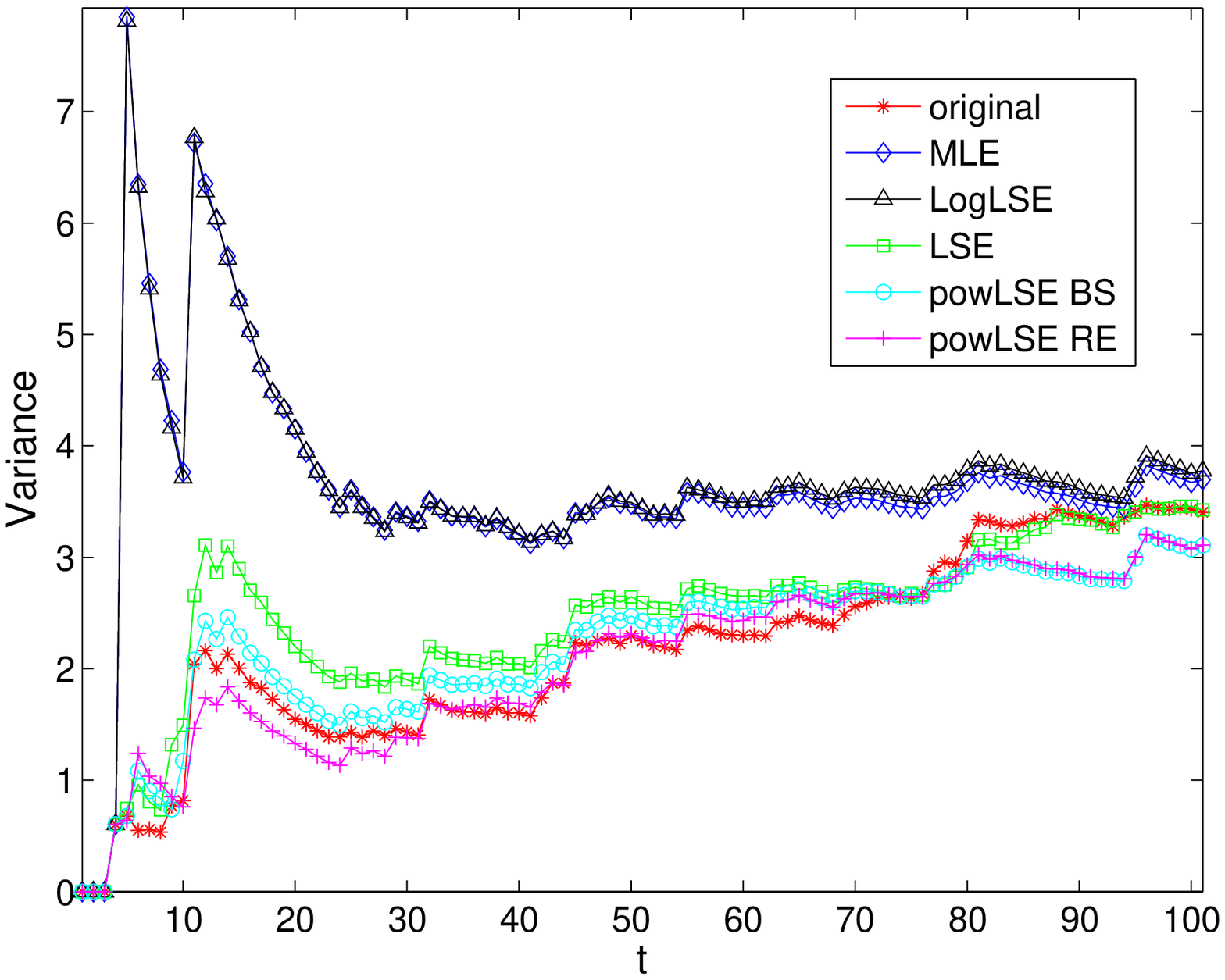}
\caption{Variances of JDM--IV with MLE, LogLSE and powLSE.}
\end{minipage}
\hspace{1cm}
\begin{minipage}{5.5cm}
\includegraphics[width=5.5cm]{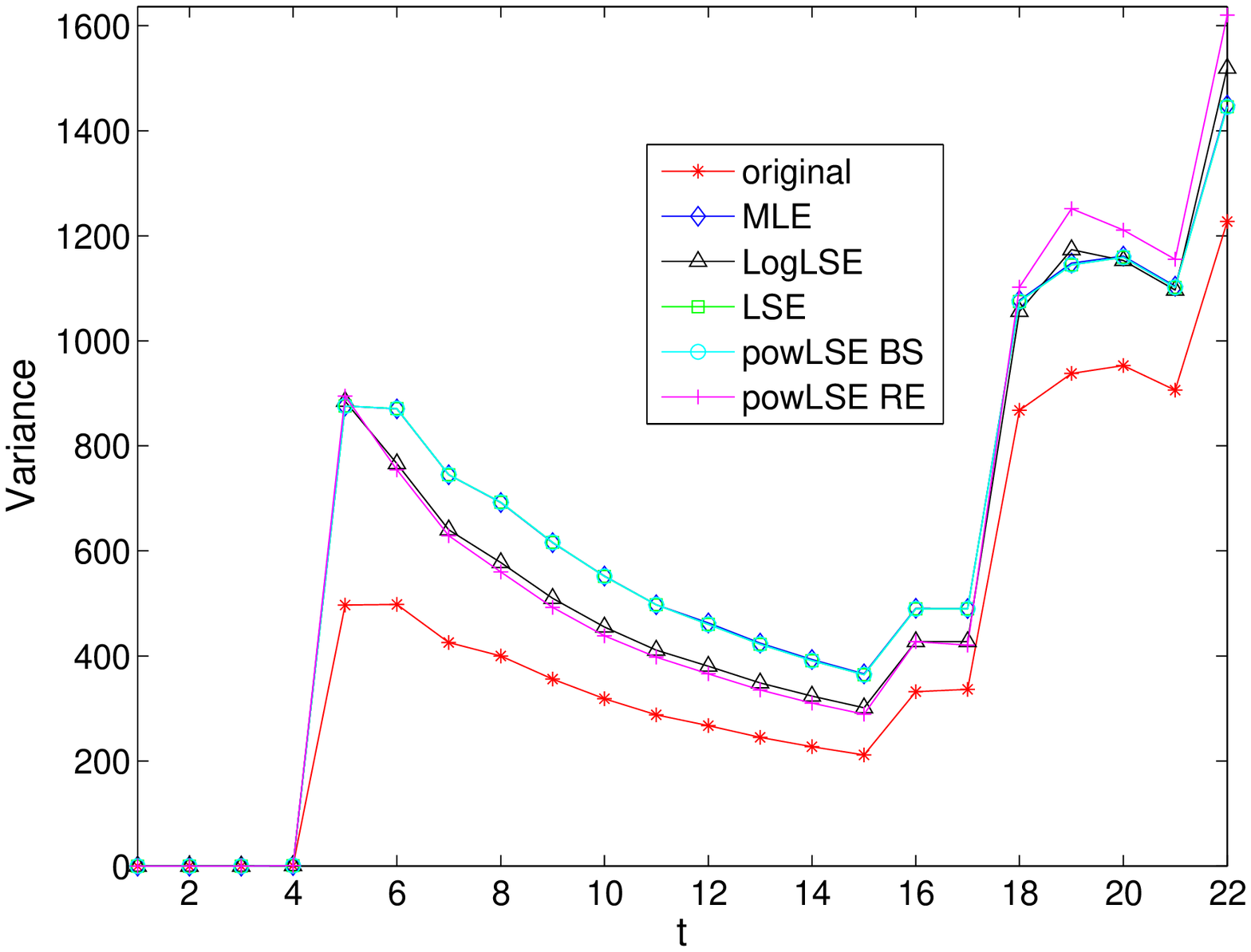}
\caption{Variances of AT\& T with MLE, LogLSE and powLSE.}
\end{minipage}
\end{center}
\end{figure}

We can conclude from Fig. 25 to Fig. 30 that all of the six failure data have heteroscedasticity along the time,
the JDM--II and JDM--IV, which have relatively small variance values, have small predictive RE values (see Table 7),
and the variance change points reflect the role of sample data affecting the deviation at the same time.
And, the heteroscedasticity provides the information of data sensitivity to modeling software reliability. All of the information would provide help for statistical modeling and explanation for bad performance.

\section{Conclusion}

A FNLSE framework is proposed, two of special cases, LogLSE and powLSE, are applied to the parameter estimation of Jelinski--Moranda model. It extends the LSE and LogLSE in software reliability and possesses the data compressing role with the proper selection of transformation function, and it is proved as a weighted LSE. Our motivation
and modeling procedure are different from famous Box--Cox transformation. It is also treated as a general model to
discuss statistical data analysis with non--normality and heteroscedasticity.
Furthermore, the LogLSE and powLSE of Jelinski--Moranda model are discussed and derived. Their prediction accuracies are evaluated by two statistical indexes relative error and Braun statistic. The simulation results demonstrate that both powLSE and LogLSE outperform MLE and LSE of Jelinski--Moranda model, and powLSE outperforms LogLSE with the optimal indexes. The future work will focus on the simulation evaluation on more failure data sets with FNLSE,  compare the performance of FNLSE with time--dependent model, and apply the FNLSE to the other software reliability models to generalize the estimation algorithm modeled by LSE.

\section{Acknowledgements}

The research was partially supported by 863 Project of China (2008AA02Z306), Beihang SRSA Program under grant No. 2006-49-8-4, and NSFC(10801019),

\end{document}